\documentclass[fleqn,usenatbib]{mnras}

\usepackage{txfonts}
\usepackage{pdflscape}
\usepackage[T1]{fontenc}
\usepackage{ae,aecompl}

\usepackage{color}
\usepackage{xcolor}

\usepackage{graphicx}	
\usepackage[normalem]{ulem}

\newcommand{\NII}{[N~{\sc ii}]~$\lambda$6584}
\newcommand{\ha}{H$\alpha$}
\newcommand{\SII}{[S~{\sc ii}]~$\lambda\lambda$6717,6731} 
\newcommand \vhel{\ifmmode{~V_{{\rm HEL}}}\else{~$V_{{\rm HEL}}$}\fi}
\newcommand \kms{km s$^{-1}$}
 
\title[IC~4776 and its binary central star]{The planetary nebula IC~4776 and its post-common-envelope binary central star}

\author[Paulina Sowicka et al.]{Paulina 
Sowicka,$^{1}$\thanks{E-mail: paula@camk.edu.pl} David Jones,$^{2,3}$ Romano L.~M. Corradi,$^{2,6}$ Roger Wesson,$^{4}$ 
\newauthor  Jorge Garc\'ia-Rojas,$^{2,3}$ Miguel Santander-Garc\'ia,$^{5}$   Henri M.~J. Boffin,$^{7}$ and
\newauthor Pablo Rodr\'iguez-Gil$^{2,3}$
\\
$^{1}$Nicolaus Copernicus Astronomical Center, Bartycka 18, PL-00-716 Warsaw, Poland\\
$^{2}$Instituto de Astrof\'isica de Canarias, E-38205 La Laguna, Tenerife, Spain\\
$^{3}$Departamento de Astrof\'isica, Universidad de La Laguna, E-38206 La Laguna, Tenerife, Spain\\
$^{4}$Department of Physics and Astronomy, University College London, Gower Street, London WC1E 6BT\\
$^{5}$Observatorio Astron\'omico Nacional (OAN-IGN), C/ Alfonso XII, 3, 28014, Madrid, Spain\\
$^{6}$GRANTECAN, Cuesta de San Jos\'e s/n, E-38712 , Bre\~na Baja, La Palma, Spain\\
$^{7}$European Southern Observatory, Karl Schwarzschild Strasse 2, 85748 Garching, Germany}

\date{Accepted XXX. Received YYY; in original form ZZZ}

\pubyear{2017}

\begin{document}
\label{firstpage}
\pagerange{\pageref{firstpage}--\pageref{lastpage}}
\maketitle

\begin{abstract}
We present a detailed analysis of IC~4776, a planetary nebula displaying a morphology believed to be typical of central star binarity.  The nebula is shown to comprise a compact hourglass-shaped central region and a pair of precessing jet-like structures.  Time-resolved spectroscopy of its central star reveals periodic radial velocity variability consistent with a binary system.  While the data are insufficient to accurately determine the parameters of the binary, the most likely solutions indicate that the secondary is probably a low-mass main sequence star.  An empirical analysis of the chemical abundances in IC~4776 indicates that the common-envelope phase may have cut short the AGB evolution of the progenitor.  Abundances calculated from recombination lines are found to be discrepant by a factor of approximately two relative to those calculated using collisionally excited lines, suggesting a possible correlation between low abundance discrepancy factors and intermediate-period post-common-envelope central stars and/or Wolf-Rayet central stars.  The detection of a radial velocity variability associated with binarity in the central star of IC~4776 may be indicative of a significant population of (intermediate-period) post-common-envelope binary central stars which would be undetected by classic photometric monitoring techniques. 
\end{abstract}




\section{Introduction}

Planetary nebulae (PNe) are the intricate, glowing shells of gas ejected by low- and intermediate- mass stars at the end of their asymptotic giant branch (AGB) evolution which are then ionized by the emerging pre-white dwarf core. With some $\sim$80 per cent of all PNe showing deviation from spherical symmetry \citep{parker06}, it has proven impossible to understand their structures in terms of single star evolution \citep{soker06,nordhaus07,garcia-segura14}, with binarity frequently invoked to explain their diverse, often strongly axisymmetrical morphologies \citep{demarco09,jones17c}. 

While a lower limit to the close-binary central star fraction is well constrained (at $\sim$20\%) by photometric monitoring surveys \citep{miszalski09a}, it is insufficient to explain all aspherical PNe. The remaining aspherical PNe are generally understood to be the products of mergers, wider binaries and/or weaker binary interactions \citep[i.e. the engulfment of a Jovian mass planet; ][]{demarco11}. This hypothesis is supported by common-envelope (CE) population synthesis models, which predict a significant number of post-CE binaries with orbital periods of several days to a few weeks \citep[e.g.][]{han95} while almost all of the known post-CE central stars have periods less than one day \citep{jones17c}. 

The lack of known post-CE central stars with intermediate periods \citep[only three are known to have periods greater than three days,][]{mendez81,manick15,miszalski17} is, perhaps, not unreasonable given that the majority of the effort has been focused on photometric monitoring which becomes particularly insensitive at these longer periods \citep{demarco08,jones17c}. Recent work, however, has shown that targeted radial velocity monitoring can begin to reveal these missing binary systems \citep{miszalski17,jones17b}. Constraining this population is of particular interest given that observations of  ``naked'' (i.e. those with no surrounding PN) white dwarf plus main sequence binaries find a similar dearth of intermediate period systems \citep{nebot11}, strongly indicating that the lack of known systems is not purely an observational bias. Understanding to what extent this population is truly absent (rather than just difficult to detect) will greatly further our understanding of the common envelope process itself \citep{toonen13}.

\begin{figure*}
    \centering
	\includegraphics[scale=0.60]{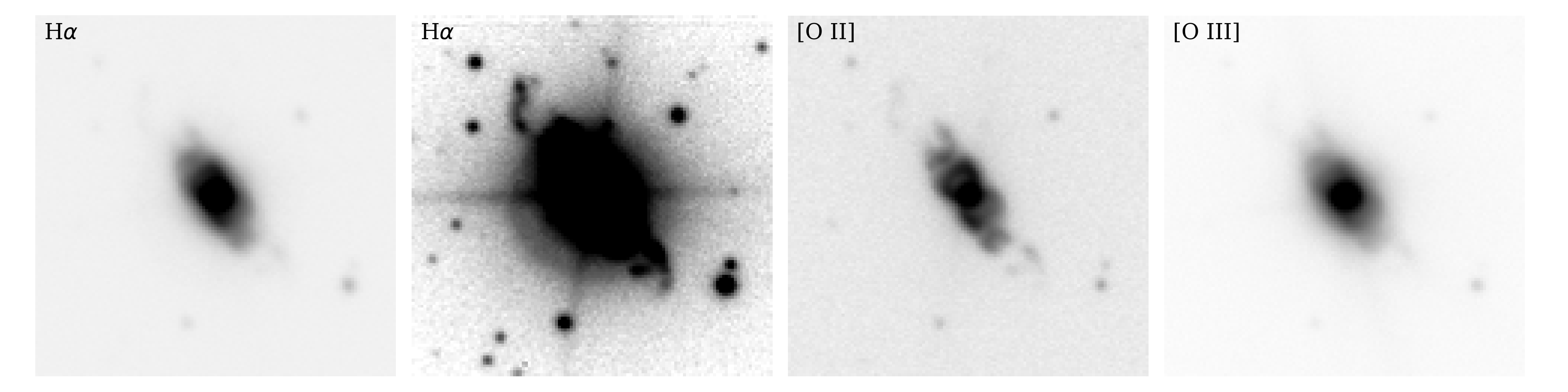}
    \caption{FORS2 narrowband imagery of IC~4776.  Both H$\alpha$ images include the light of [N~\textsc{ii}], and the central region of the rightmost H$\alpha$ image is heavily saturated but most clearly shows the structure of the faint, precessing jets.  All images measure 0.5\arcmin{}$\times$0.5\arcmin{}, north is up and east is left.}
    \label{fig:narrowband_images}
\end{figure*}

IC~4776 is a relatively bright, small planetary nebula comprising a central, bipolar, hourglass-like structure and an extended, jet-like structure revealed for the first time by the images presented in this paper (see Figure~\ref{fig:narrowband_images} and Section \ref{subsec:nebims}).   Both jets and hourglass structures have been strongly linked to binary central stars \citep{miszalski09b}.  The nebula has been shown to display dual-dust chemistry \citep[i.e.\ features associated with both carbon- and oxygen-rich dust;][]{perea09,gorny10}, which has also been linked to a possible binary evolution of the nebular progenitor \citep{guzman14}. The spectral type of its central star is unclear but has been classified by various authors as [WC] \citep[e.g.\ ][]{aller85}.  The apparent presence of narrow emission lines has also earned it a \textit{wels} classification \citep{tylenda93}, now generally not considered a valid classification given that in many cases the emission lines themselves do not originate from the stars but from their host nebulae \citep{basurah16} or are the product of the irradiation of a main-sequence companion \citep{miszalski11b}.  

Based on the likelihood that it hosts a binary central star, IC~4776 was selected for further study as part of a programme to search for binary central stars through time-resolved radial velocity monitoring.  Here, we present the results of that radial velocity study, revealing variability consistent with an intermediate period, post-CE binary.  We furthermore present spatio-kinematic and chemical analyses of the nebula itself, in order to further constrain the relationship between the nebula and the probable central binary.


\section{Nebular morphology and kinematics}
\label{sec:neb}

\subsection{FORS2 narrowband imagery}
\label{subsec:nebims}

Narrowband imagery of IC~4776 was obtained using the FORS2 instrument mounted on ESO's VLT-UT1 \citep{FORS}.
Exposures were acquired in the following emission lines: \ha{}+[N~\textsc{ii}] (5s exposure time on 2012 September 5, 30s on 2016 June 11),  [O~\textsc{ii}]~$\lambda$3727 (20s 2012 September 3), and [O~\textsc{iii}]~$\lambda$5007 (5s 2012 September 5).  In each case the seeing was better than 1\arcsec{}. The debiased and flat-fielded images are presented in Figure~\ref{fig:narrowband_images}.

In all three bands, the nebula shows a similar morphology, namely an ``X"-shape, which is most prominently visible in the light of [O~\textsc{ii}]~$\lambda$3727. At all three wavelengths the central shell appears the brightest, while only at [O~\textsc{ii}]~$\lambda$3727 it is possible to discern its detailed structure. This hourglass morphology is extremely similar to that of MyCn~18, a PN often hypothesised to have originated from a binary interaction owing to its similarity to some nebulae surrounding symbiotic binary stars \citep{corradi93}.  MyCn~18, in addition to its central hourglass, displays a system of high velocity knots which may well be analogous to the jets observed in IC~4776 \citep{bryce97,oconnor00}, further highlighting the similarities between the two nebulae. The jets of IC~4776 are evident in all filters but most prominently in the light of \ha{} where their remarkable structure is clearest.  The evident curvature of the jets is a strong indication of precession, which can be a natural product of a central binary system \citep{raga09}.  Such precessing jets have been observed in several close-binary PNe, e.g.\ Fg~1 \citep{boffin12b}, ETHOS 1 \citep{miszalski11a}, and NGC~6337 \citep{garcia-diaz09,hillwig16}, providing a solid observational link between jets and binarity.

\subsection{FLAMES-GIRAFFE integral field spectroscopy}
\label{subsec:nebspec}

\begin{figure}
	\includegraphics[scale=0.55]{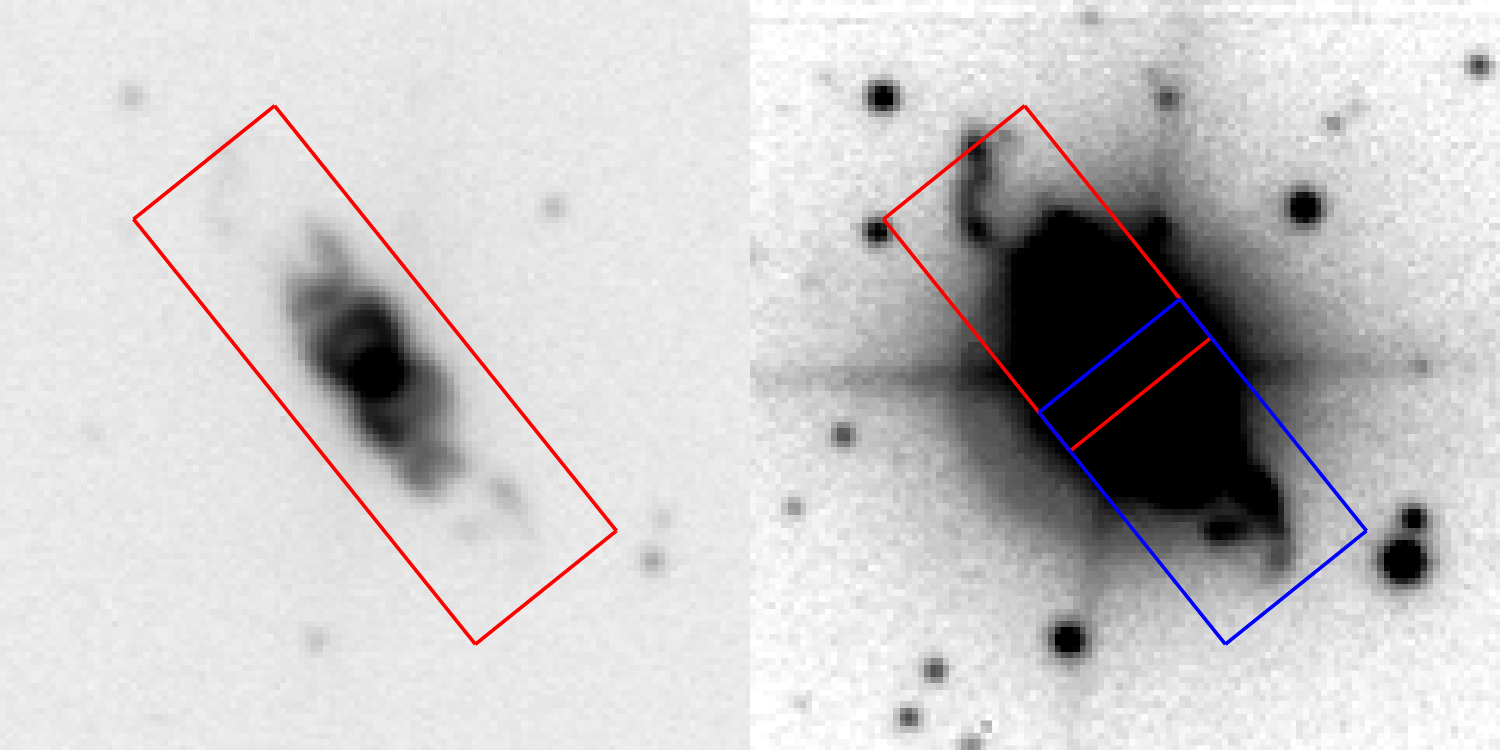}
    \caption{The ARGUS-IFU coverage of IC 4776. Each image measures 1\arcmin{}$\times$ 1\arcmin{}. North is up, East to the left. \textit{Left:} the two pointings merged together, overplotted on FORS2 image in the light of [O II]. \textit{Right:} the two pointings shown separately, overplotted on FORS2 image in the light of \ha{} in order to highlight the spatial coverage of the jets.
    }
    \label{fig:Flames_pointings}
\end{figure}

Spatially-resolved, high-resolution, integral-field spectroscopy of IC~4776 was obtained on August 8 2013 using FLAMES-GIRAFFE mounted on ESO's VLT-UT2 \citep{FLAMES}.  The seeing during the observations did not exceed 1\arcsec{}. The ARGUS integral field unit, employed in its standard `1:1' scale mode (resulting in a total aperture of 11.5 x 7.3\arcsec{} sampled by an array of 22$\times$14 0.52\arcsec{}$\times$0.52\arcsec{} microlenses), was used to feed the GIRAFFE spectrograph setup with the H665.0/HR15 grating (providing a resolution of R$\sim$30\,000 in the range 6470\AA{} $\lesssim\lambda\lesssim$ 6790\AA{}). ARGUS was oriented at P.A. = 38$^{\circ}$ (to coincide with the symmetry axis of the nebula) and with only two telescope pointings, offset by about 10\arcsec{} to each other, we were able to obtain a complete coverage of the nebula (which is $\sim$20\arcsec{} long including the faint extended emission), as shown in Figure~\ref{fig:Flames_pointings}. Two exposures of 245-s were obtained for each of the two pointings, with the resulting data reduced and combined  to produce a single, wavelength-calibrated data-cube for each pointing using the standard ESO-GIRAFFE pipeline.  Finally, the data from the two pointings were corrected to heliocentric velocity and combined to form a single data-cube using specially written \textsc{python} routines.

In the wavelength range of the ARGUS data there are several spectral lines: \ha, [N~\textsc{ii}]~$\lambda\lambda$6548,6584 and the \SII{} doublet.  In order to probe the kinematics and morphology of the nebula, we focus on the \NII{} line given that it is brighter than \SII{} and has a lower thermal width than \ha{}.  

\begin{figure*}
	\centering
    \includegraphics[scale=0.3]{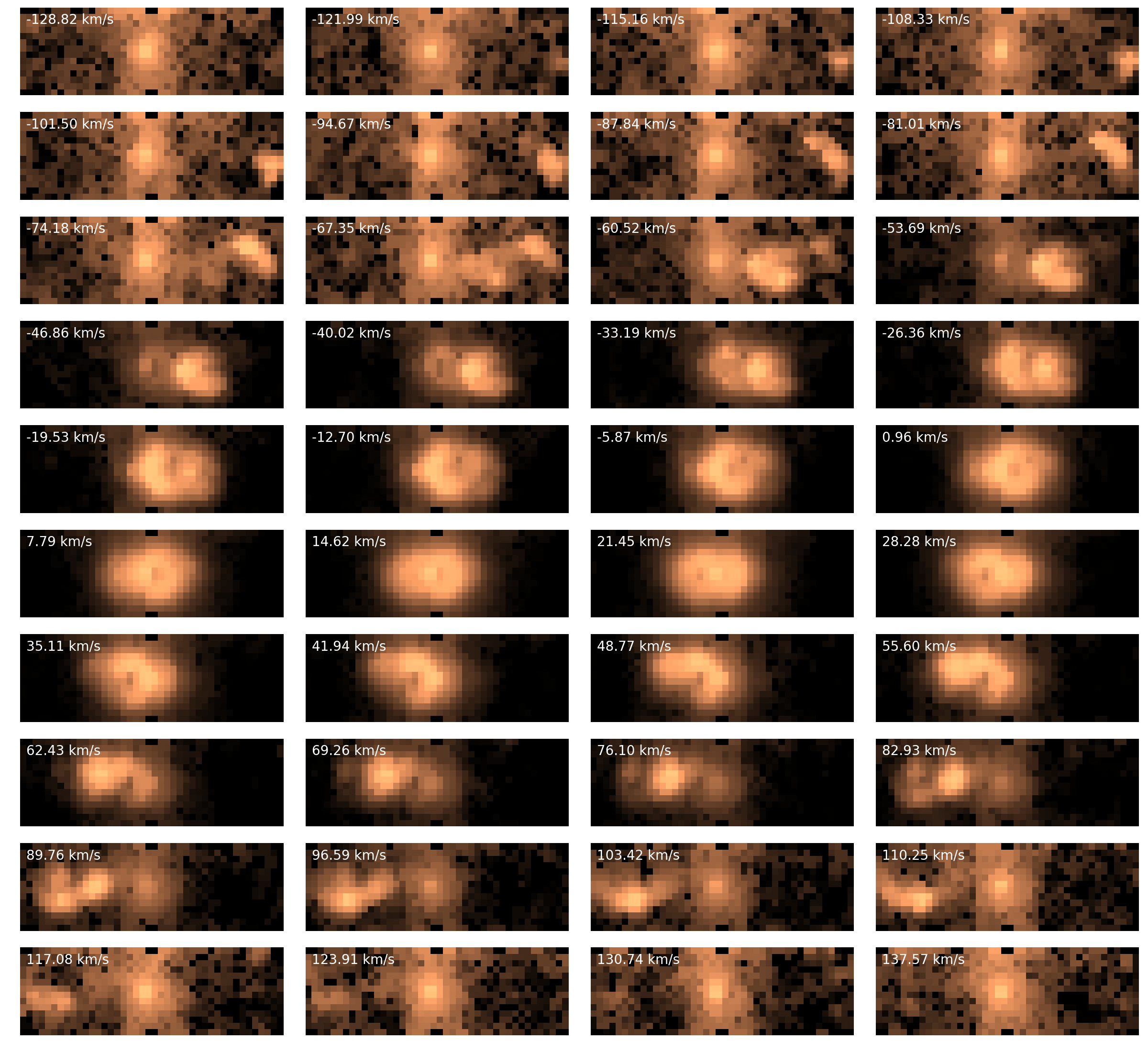}
    \caption{Velocity channel map for the [N~\textsc{ii}]~$\lambda$6583.45 emission from IC~4776.  Velocities are corrected to Heliocentric, and each channel represents the summed emission of three ``slices'' of the reduced data cube, corresponding to a velocity bin of 6.83 km~s$^{-1}$.  The logarithmic display limits for each channel are calculated individually in order to maximise the features visible on each map.}
    \label{fig:velocity_slices}
\end{figure*}


Channel maps for the \NII{} line are shown in Figure~\ref{fig:velocity_slices}.  Each channel shows a cross-section through the nebula at a different velocity relative to the rest velocity of \NII{}, starting from the top left-hand side channel being the most blue-shifted (velocity of about $-130$ km~s$^{-1}$), ending on the bottom right-hand side channel being the most red-shifted (velocity of about 137 km~s$^{-1}$). Each channel is a sum of three slices of the reduced cube where the wavelength step between slices is approximately 0.05\AA{} (i.e.,\ each channel represents a range of 0.15\AA{}$\equiv$6.83 km~s$^{-1}$ at \NII{}). The brightness/contrast is calculated individually for each channel using a logarithmic function in order to display the maximum amount of structural information possible. 

The systemic velocity of the nebula has been previously found to be $v_{sys}$=16.3$\pm$0.6 \kms{} \citep{durand98}, and around this velocity the channel maps present two circular structures joined at the position of the central star.  This can be interpreted as the contribution of the two lobes of an inclined bipolar shell, as inferred from the imagery presented in Section \ref{sec:neb}.  Further from this central velocity, one lobe begins to dominate, just as would be expected from an inclined hourglass.  Yet further still from the systemic velocity (e.g.\ $v_{sys}\pm\sim$20\kms{}), a second component begins to contribute significantly to the observed emission in each channel map, initially appearing close to the central star and then moving away at velocities furthest from the systemic velocity.  This emission feature clearly originates from the jets, presenting with a similarly curved appearance at the largest velocities.  The jets themselves also present clearly greater velocities than the central shell (the maximum extent of which is at roughly $v_{sys}\pm$35\kms{}), reaching more than 100\kms{} away from the systemic velocity.  Furthermore, there is an evident asymmetry in the jet velocities given that the blue-shifted jet reaches roughly $v_{sys}-$130\kms{}, while the red-shifted component only reaches as far as $v_{sys}+$100\kms{}.  Figure \ref{fig:Flames_pointings} reinforces this idea as in the high contrast image the Northern jet extends beyond the region encompassed by the FLAMES-GIRAFFE observations (centred on the central star), while the Southern jet is apparently well covered.  Given the clear precession visible in both the imagery and IFU spectroscopy it is not possible to derive a simple correlation between line-of-sight velocity and distance from the central star which would further constrain the asymmetry.

\begin{figure}
	\centering
   \includegraphics[scale=0.5]{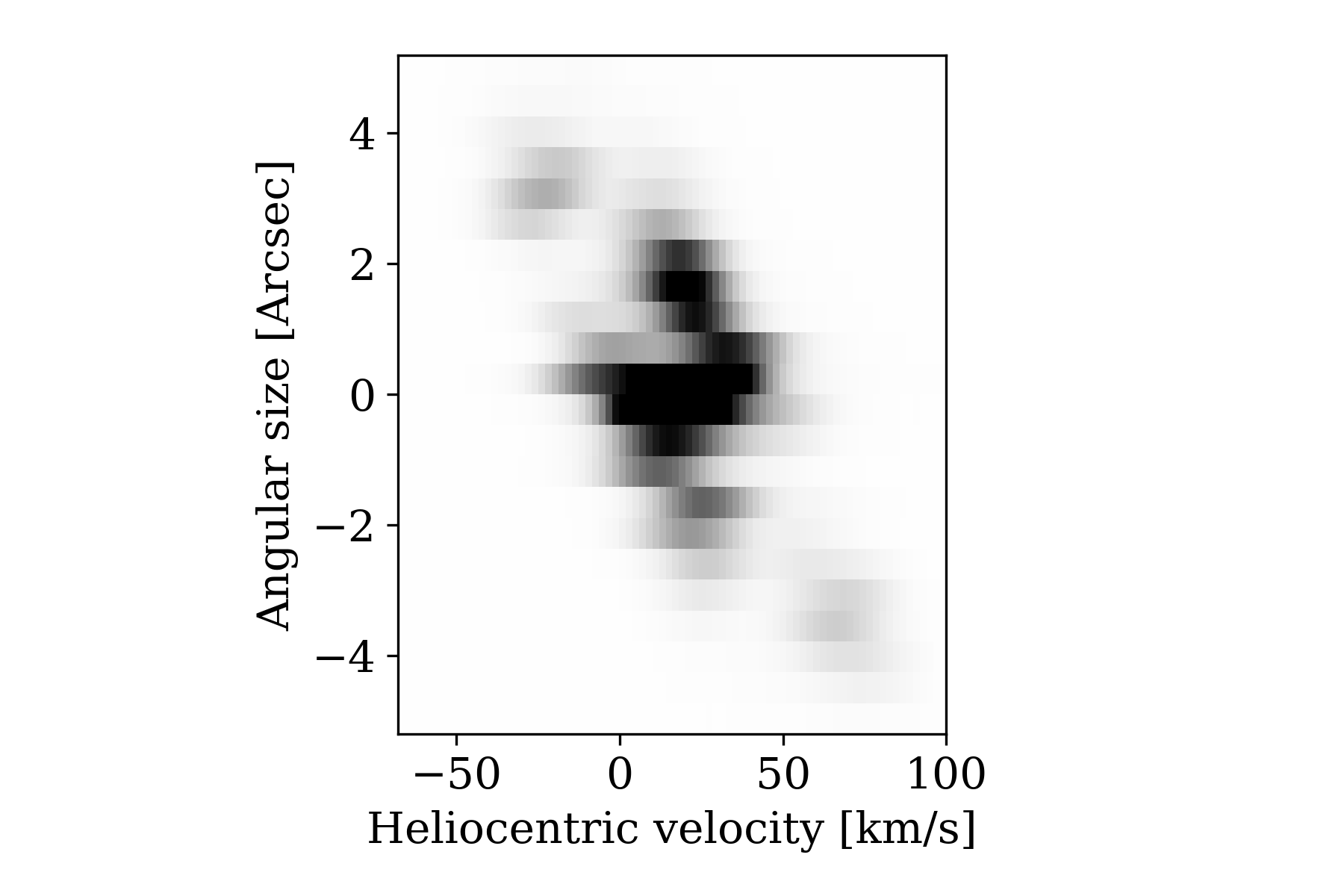}
   \caption{Simulated position velocity corresponding to an [N~\textsc{ii}]~$\lambda$6583.45\AA{} longslit, with width $\sim$1\arcsec{}, placed along the symmetry axis of the nebula.}
   \label{fig:longslit}
\end{figure}

A 2 spaxel region surrounding the central star was collapsed (equivalent to a slit-width of 1.04\arcsec{}) in order to produce a simulated longslit spectrum. The position-velocity array of this simulated longslit is shown in Figure~\ref{fig:longslit}.  The position-velocity array shows a clear ``X''-shape, typical of such bipolar, hourglass structures, as well as end-caps corresponding to the brightest regions of jet emission (further emission from the jets can be seen with increased contrast which blows out the central nebula).

\begin{figure}
   \includegraphics[width=\columnwidth]{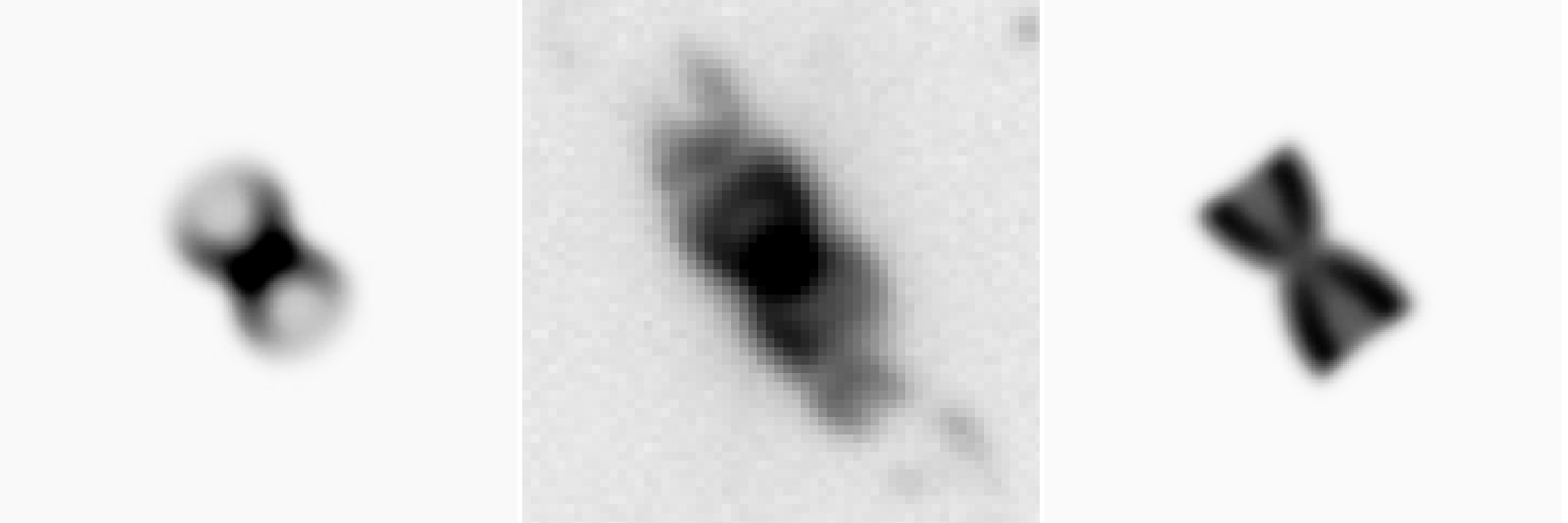}
   \caption{Synthetic images of the \textsc{shape} model of IC~4776 at the derived inclination of 42\degr{} (left) and at an inclination of 90\degr{} (right).  The central image is the FORS2 [O~\textsc{ii}] image (reproduced from  Fig.\ \ref{fig:narrowband_images} for direct comparison).  Each image measures 0.25\arcmin{}$\times$0.25\arcmin{}.}
   \label{fig:shapemodel}
\end{figure}

\subsection{Shape modelling}
\label{sec:nebshape}

A spatio-kinematic model of the central hourglass of IC~4776 was constructed based on the narrowband imagery and spatially-resolved spectroscopy presented above.  The \textsc{shape} software was used, following the standard workflow presented in \citet{steffen10}, varying both morphological and kinematic parameters over a wide-range of parameter space with the best fit being determined from a by-eye comparison of synthetic images and velocity channel maps to the observations. The jets themselves are too complex to be treated by a simple spatio-kinematic model, and their detailed modelling is reserved for a subsequent paper (Santander-Garc\'ia et al., in prep). Although, it is worth noting that, in principle, the data are of sufficient quality to derive the precession properties of the jets, which can then be related to the parameters of the central binary star \citep[e.g.][]{raga09,boffin12b} at the time the jet was launched (constrained by the kinematical age of the jet).

The best-fitting \textsc{shape} model, as expected from appearance of the nebula in the narrowband imagery, comprises an hourglass-like structure whose symmetry axis is inclined at 42\degr{}$\pm$4\degr{} with respect to the plane of the sky (see Figure~\ref{fig:shapemodel}).  The velocity structure of the nebula is found to be well represented by a flow-law whereby all velocities are radial from the central star and proportional to the distance from the central star. Such flow-laws, often referred to as ``Hubble flows'', are generally taken to imply an eruptive event in which the majority of the nebular shaping occurred during a brief time period.  However, \cite{steffen09} showed that while deviations from a ``Hubble flow'' should be appreciable and easily observed in PNe formed via the classical interacting stellar winds model \citep{kwok78,kahn85}, such a simple flow law will provide a good first order estimation of the overall morphological and kinematic properties of the nebula.  Particularly given the relatively low spatial resolution of the FLAMES-GIRAFFE observations c.f.\ the size of IC~4776, we are unable to constrain the presence (or not) of such deviations from a ``Hubble flow''.

The systemic velocity of the nebula is found to agree well with the literature value of 16.3$\pm$0.6 km~s$^{-1}$ \citep{durand98}, and was fixed in the modelling. The kinematical age of the nebula is found to be approximately 300 yr kpc$^{-1}$ (with a significant uncertainty due to the small size of the nebula).  While there is significant variation in distance determinations for IC~4776 in the literature, the kinematical age would imply a rather young nebula at all but the very largest distances.  The distance of 1.01 kpc determined by \citet{phillips84} would seem to imply an impossibly young nebula,  while the distances derived by \citet[4.97 kpc]{stanghellini08} and \citet[4.44$\pm$1.27 kpc]{frew16} would give a more reasonable age of $\sim$1500 years.

\section{Nebular chemistry}
 
 High-resolution longslit spectroscopy of the nebula IC~4776 was acquired on August 9 2016 using the Ultra-violet and Visual Echelle Spectrograph (UVES) mounted on ESO's VLT-UT2 \citep{dekker00}. Exposures were acquired using both arms of the spectrograph with the following set-ups. Using Dichroic 1 (3$\times$ 150s exposures, 1$\times$ 20s exposure), the blue-arm was set to a central wavelength of 3460\AA{} (with the standard HER\_5 blocking filter) while the red-arm was set to a central wavelength of 5800\AA{} (with the standard SHP700 blocking filter).  Using Dichroic 2 (3$\times$ 300s exposures, 1$\times$ 60s exposure, 1$\times$ 20s exposure), the blue-arm was set to a central wavelength of 4370\AA{} (again with the standard HER\_5 blocking filter) while the red-arm was set to 8600\AA{} (with the standard OG590 blocking filter).  A slit-width of 2.4\arcsec{} was employed, and the seeing during the observations was approximately 1.5\arcsec{}. The spectra were reduced using the standard UVES pipeline and flux-calibrated using 300s exposures of the standard star LTT~7987 taken with the same set-ups directly following the observations of IC~4776.  Collectively, the spectra provide near-continuous wavelength coverage from $\sim$3000--10\,000\AA{} (see Figure~\ref{fig:alfaneatfull}).  To construct the final spectrum, we took the median of the flux at each wavelength from the longest exposures in each setup, and where lines were saturated, replaced the affected ranges with values from the shorter exposures in which saturation was not an issue.
 
\begin{figure*}
     \centering
   \includegraphics[width=\textwidth]{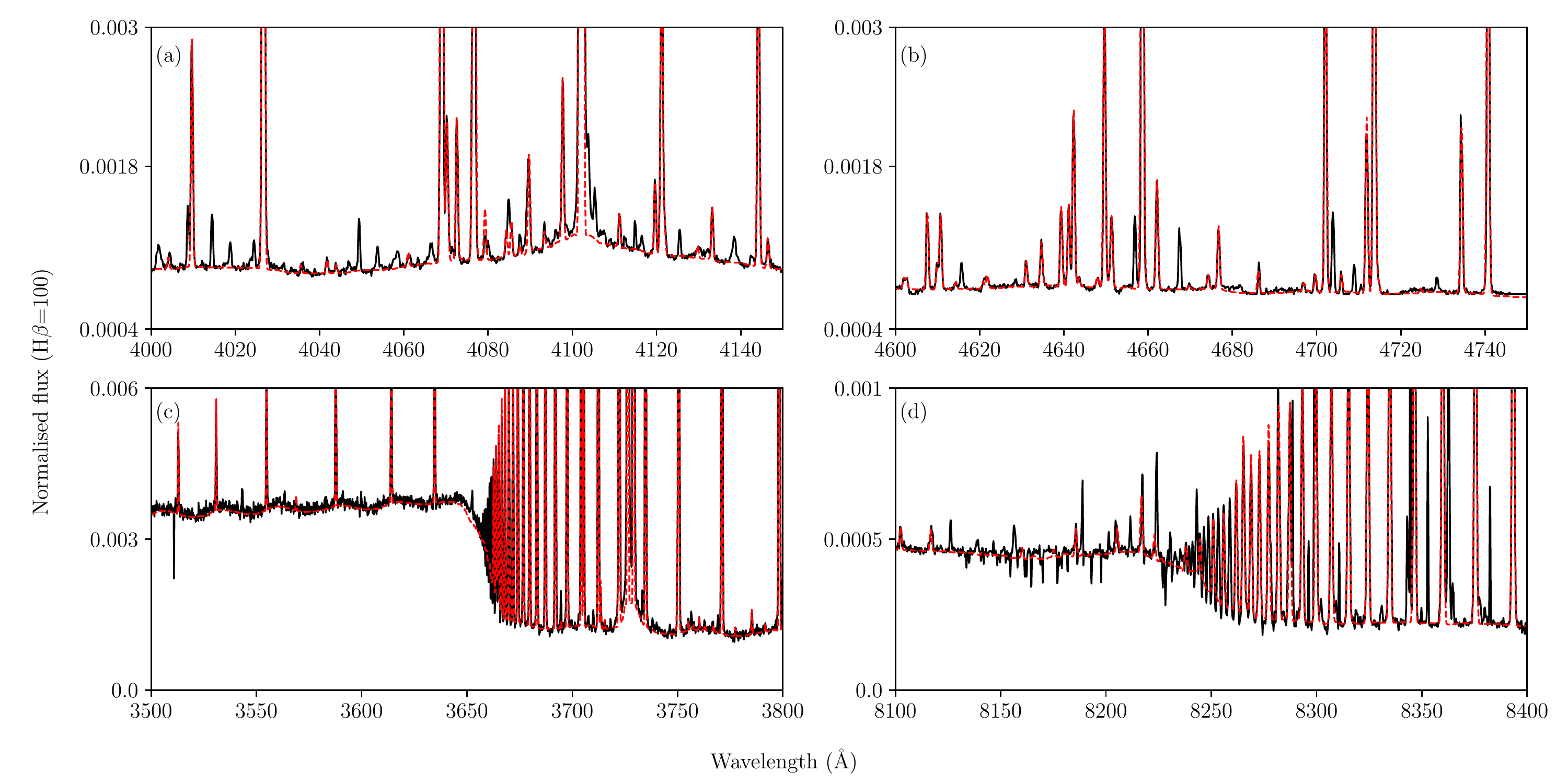}
   \caption{The \textsc{alfa} spectral fit (red-dashed) overlaid on top of the combined, observed UVES spectra (black-solid).  The panels highlight key regions used in the \textsc{neat} analysis: (a) and (b) show two regions containing O~\textsc{ii} recombination lines critical for deriving the nebular adf, while the panels (c) and (d) cover the Balmer and Paschen jumps, respectively.  The full spectrum is shown in figure \ref{fig:alfaneatfull}.}
   \label{fig:alfaneat}
\end{figure*}

 
 Emission line fluxes were measured using the \textsc{alfa} code \citep{wesson16}, which optimises parameter fits to the line profiles using a genetic algorithm following the subtraction of a global continuum.  The effectiveness of \textsc{alfa} in measuring PN line fluxes has previously been demonstrated in several papers \citep[e.g.][]{jones16b}, and is further demonstrated in Figures~\ref{fig:alfaneat} and \ref{fig:alfaneatfull}.  \textsc{alfa} assumes that all lines have a Gaussian profile.  At the high resolution of the UVES spectra, velocity structure is evident, and lines of different ionisation have different profiles.  For the purposes of chemical analysis, we binned the spectra by a factor of 10 in wavelength, resulting in Gaussian profiles for all lines.

353 emission lines were measured, of which 329 were resolved and 24 were blends of two or more lines.  \'Echelle spectra may contain many spurious features due to bleeding of strong lines from adjacent orders.  In general, these do not affect the analysis, as ALFA only attempts to fit features close to the wavelengths of known emission lines.  Spurious features not fitted by ALFA may be seen in panels (a) and (b) of Figure~\ref{fig:alfaneat} (though they are also present in all panels of Figure~\ref{fig:alfaneatfull}).  However, in a few cases, the bleeding may be blended with nebular emission. The O~\textsc{ii} recombination line at 4089{\AA} is one such case, and this impacts the estimate of the temperature and density from the ratios of O~\textsc{ii} lines, as discussed below.  Table \ref{tab:fluxes} lists the observed and dereddened fluxes measured for all emission lines measured in the spectra, together with their 1$\sigma$ uncertainties. 

Physical conditions as well as chemical abundances were calculated from the emission line fluxes using the \textsc{neat} code \citep{wesson12}.  The code uses Monte Carlo techniques to propagate uncertainties in line flux measurements into the derived quantities \citep[for full details please see][]{wesson12}.  The physical parameters determined for IC~4776 are listed in Table \ref{tab:phys}.

 \begin{table}
	\centering
	\caption{Extinction, temperatures and densities of IC~4776 as derived using the \textsc{neat} code, as well as the temperatures and densities used for the determination of the chemical abundances listed in Tables \ref{tab:ionic} and \ref{tab:abund}.}
	\label{tab:phys}
	\begin{tabular}{rrl}
		\hline
        c(H$\beta)$ = & ${  0.22}$ &$\pm0.03$ \\
		$n_e$([O~{\sc ii}]) (cm$^{-3}$) = & ${ 9660}$&$^{+10020}_{-3610}$ \\
		$n_e$([S~{\sc ii}]) (cm$^{-3}$) = & ${11600}$&$^{+ 9000}_{-3600}$ \\
		$n_e$([Cl~{\sc iii}]) (cm$^{-3}$) = & ${18600}$&$^{+ 5500}_{-3900}$ \\
		$n_e$([Ar~{\sc iv}]) (cm$^{-3}$) = & ${41300}$&$^{+ 4500}_{-4000}$ \\
        $n_e$(BJ) (cm$^{-3}$) = & ${ 11400}$&$^{+ 16900}_{ -6800}$ \\
        $n_e$(PJ) (cm$^{-3}$) = & ${ 57500}$&$^{+ 54500}_{-28000}$ \\
		$T_e$([O~{\sc ii}]) (K) = & ${30800}$&$^{+ 4200}_{-16800}$ \\
		$T_e$([S~{\sc ii}]) (K) = & ${18000}$&$^{+17000}_{-9300}$ \\
		$T_e$([N~{\sc ii}]) (K) = & ${12000}$&$^{+  1000}_{-1200}$ \\
		$T_e$([O~{\sc iii}]) (K) = & ${10000}$&$\pm{  200}$ \\
		$T_e$([Ar~{\sc iii}]) (K) = & ${ 9070}$&$\pm{  180}$ \\
		$T_e$([S~{\sc iii}]) (K) = & ${11100}$&$\pm{  400}$ \\
		$T_e$(BJ) (K) = & ${ 8920}$&$^{+  760}_{ -700}$ \\
		$T_e$(PJ) (K) = & ${ 7240}$&$^{+ 1070}_{ -940}$ \\
		$T_e$(He 5876/4471) (K) = & ${ 3180}$&$^{+ 1060}_{ -800}$ \\
		$T_e$(He 6678/4471) (K) = & ${ 3970}$&$^{+ 1710}_{-1190}$ \\
		$T_e$(O~\textsc{ii} ORLs) (K) = & ${ 3560}$&$^{+ 3710}_{-1960}$ \\
		$n_e$(O~\textsc{ii} ORLs) (cm$^{-3}$) = & ${ 2850}$&$^{+ 1430}_{-1240}$ \\
        \hline
\end{tabular}
\end{table}

The extinction towards the nebula was estimated from the flux-weighted average of the ratios of H$\alpha$, H$\gamma$ and H$\delta$ to H$\beta$, and the Galactic extinction law of \cite{howarth83}.  H$\gamma$ and H$\delta$ are in the same spectrum as H$\beta$, while H$\alpha$ is in a separate spectrum.  The three lines give consistent estimates of the logarithmic extinction at H$\beta$, c(H$\beta$), which is 0.22$\pm$0.03. 

\begin{figure}
     \centering
   \includegraphics[width=\columnwidth]{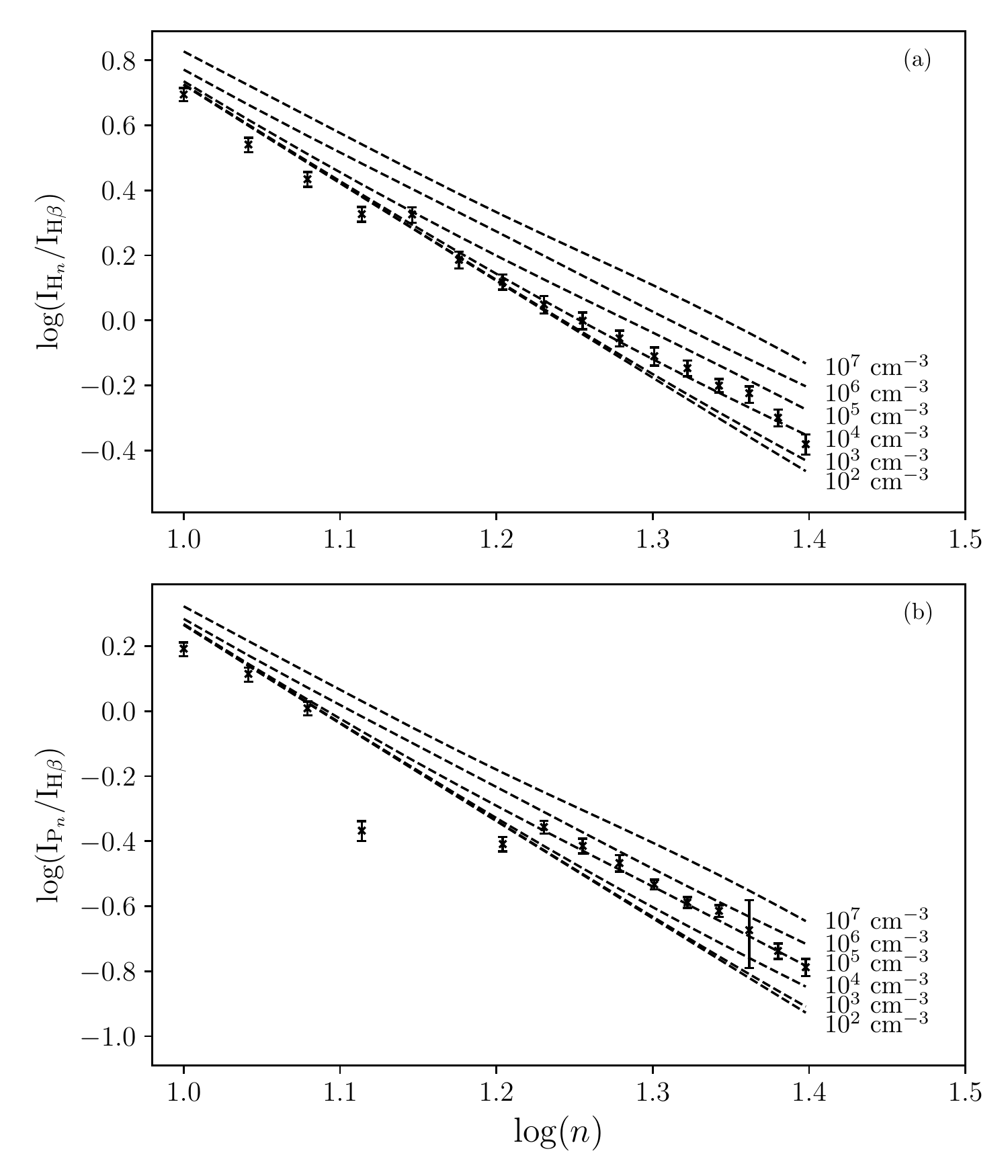}
   \caption{Observed intensities (relative to H$\beta$=100) of high-order Balmer (a) and Paschen (b) lines as a function of the principal quantum number $n$. The dashed curves show the predicted intensities of the lines as a function of electron density, at a temperature of $10\,000$K as derived from the [O~{\sc iii}] nebular to auroral line ratio.  The plots highlight the general tendency towards larger densities ($\sim$10$^4$--10$^5$cm$^{-3}$), consistent with those measured from CEL diagnostics.  Note that H14 is blended with a line of [S~{\sc iii}] and, as such, its flux is overestimated. P13 and P16 lie close to the ends of two of our spectra where the flux calibration uncertainties are large (and not reflected in the plotted uncertainties which are simply the statistical uncertainties on the flux measurement), and as such their fluxes are almost certainly underestimated given that they fall well below the trend of the other measured lines.  P14 and P15 are not shown as they lie in a gap in our spectral coverage.}
   \label{fig:balmber}
\end{figure}

The nebular density is estimated using several standard collisionally excited line (CEL) diagnostics, as well as from the Balmer and Paschen decrements, and the ratios of O~\textsc{ii} recombination lines.  The CEL diagnostics give relatively high densities of 10000--30000 cm$^{-3}$; the Paschen and Balmer decrements also imply high densities, of 10$^4$--10$^5$ cm$^{-3}$, although with larger uncertainties (see Figure~\ref{fig:balmber}).  The O~\textsc{ii} recombination line ratios imply a lower density of $\sim$3000 cm$^{-3}$, with the caveat that the 4089~\AA{} line flux on which the ratios rely may be overestimated due to bleeding from adjacent orders.  If its flux is overestimated, the density derived is underestimated.

Temperatures determined from CEL ratios are around 10kK.  The Balmer and Paschen jumps are well detected and permit the derivation of temperatures from their magnitudes.  They are statistically consistent with each other, but slightly lower than the temperatures from CELs, with T(BJ)=8900$\pm$800K, and T(PJ)=7200$\pm$1000K, where T([O~\textsc{iii}])=10000$\pm$200K.  The ratios of the three strong helium lines at 4471, 5876 and 6678{\AA} are weakly temperature-sensitive, and these are well enough detected to enable their use as a diagnostic; the temperatures from the $\lambda$5876/$\lambda$4471 and the $\lambda$6678/$\lambda$4471 ratios are significantly lower than from other aforementioned methods, at 3--4kK.  Finally, O~\textsc{ii} recombination line ratios imply a temperature consistent with the helium line ratios, albeit with a large statistical uncertainty, in addition to the systematic uncertainty of the effect of bleeding, which would result in an underestimate of the temperature.

Chemical abundances were derived from both collisionally excited lines and recombination lines (ORLs), with total chemical abundances derived using the ionisation correction scheme of \cite{delgado-inglada14}.  Heavy element recombination lines are very well detected in the spectra, and as such it is possible to measure abundances for C$^{2+}$, C$^{3+}$, O$^{2+}$, N$^{2+}$, N$^{3+}$ and Ne$^{2+}$. The \textsc{neat} code employed for the analysis assumes a three-zone ionization scheme. The average of $n_e$([O~{\sc ii}]) and  $n_e$([S~{\sc ii}]) was adopted as representative of the density of low-ionisation zone (Ionisation Potential, IP $<$ 20 eV), while the average of $n_e$([Cl~{\sc iii}]) and $n_e$([Ar~{\sc iv}]) was applied for the medium ionisation zone (20 eV $<$IP$<$ 45 eV). The electron temperature obtained from [N~{\sc ii}] was used for the temperature of the low-ionisation zone, and the average of the temperatures derived using [O~{\sc iii}], [Ar~{\sc iii}], and [S~{\sc iii}] lines taken as the temperature of the medium-ionisation zone.  No lines of the high ionisation regime (IP$>$45 eV) are detected, so no high-ionisation conditions are assumed.  The medium-ionisation parameters were also employed in the derivation of abundances from ORLs.
The calculated abundances, from both ORLs and CELs, are listed in Tables \ref{tab:ionic} and \ref{tab:abund}.

 \begin{table}
	\centering
	\caption{Ionic abundance ratios, by number, relative to H$^+$, in IC~4776.}
	\label{tab:ionic}
	\begin{tabular}{rrl}
		\hline
        Ion & \multicolumn{2}{c}{$\frac{\mathrm{X}^{i+}}{\mathrm{H}^+}$}\\
        \hline
        He$^{+}$ (ORLs)                          & ${  0.10}$&$\pm{  2.00\times 10^{ -3}}$ \\
        He$^{2+}$ (ORLs)                         & ${  1.68\times 10^{ -5}}$&$^{+  4.90\times 10^{ -6}}_{ -4.80\times 10^{ -6}}$ \\
        C$^{0}$ (CELs)                           & ${  9.49\times 10^{ -9}}$&$^{+  7.73\times 10^{ -9}}_{ -3.40\times 10^{ -9}}$ \\
        C$^{2+}$ (ORLs)                          & ${  1.62\times 10^{ -4}}$&$\pm{  3.00\times 10^{ -6}}$ \\
        C$^{3+}$ (ORLs)                          & ${  5.46\times 10^{ -6}}$&$\pm{  9.70\times 10^{ -7}}$ \\
        N$^{+}$ (CELs)                        & ${  3.91\times 10^{ -6}}$&$^{+  1.72\times 10^{ -6}}_{ -8.70\times 10^{ -7}}$ \\
        N$^{2+}$ (ORLs)                          & ${  1.75\times 10^{ -4}}$&$\pm{  1.10\times 10^{ -5}}$ \\
        N$^{3+}$ (ORLs)                          & ${  1.07\times 10^{ -5}}$&$\pm{  6.00\times 10^{ -7}}$ \\
        O$^{+}$ (CELs)                           & ${  2.22\times 10^{ -5}}$&$^{+  1.77\times 10^{ -5}}_{ -6.80\times 10^{ -6}}$ \\
        O$^{2+}$ (ORLs)                         & ${  6.04\times 10^{ -4}}$&$^{+3.20\times 10^{ -5}}_{ -3.00\times 10^{ -5}}$ \\
        O$^{2+}$ (CELs)                          & ${  3.44\times 10^{ -4}}$&$^{+  2.20\times 10^{ -5}}_{ -2.00\times 10^{ -5}}$ \\
        adf (O$^{2+}$)                    	& ${  1.75}$&$^{+  0.15}_{ -0.14}$ \\
        Ne$^{2+}$ (ORLs)                         & ${  1.71\times 10^{ -4}}$&$^{+  2.20\times 10^{ -5}}_{ -1.90\times 10^{ -5}}$ \\
        Ne$^{2+}$ (CELs)                         & ${  8.91\times 10^{ -5}}$&$^{+  5.70\times 10^{ -6}}_{ -5.30\times 10^{ -6}}$ \\
        adf (Ne$^{2+}$)                   & ${  1.92}$&$^{+0.28}_{ -0.24}$ \\
        Ar$^{2+}$ (CELs)                         & ${  1.34\times 10^{ -6}}$&$^{+  1.00\times 10^{ -7}}_{ -9.00\times 10^{ -8}}$ \\
        Ar$^{3+}$ (CELs)                         & ${  1.15\times 10^{ -7}}$&$^{+  8.00\times 10^{ -9}}_{ -7.00\times 10^{ -9}}$ \\
        S$^{+}$ (CELs)                           & ${  2.60\times 10^{ -7}}$&$^{+  1.55\times 10^{ -7}}_{ -5.80\times 10^{ -8}}$ \\
        S$^{2+}$ (CELs)                          & ${  3.65\times 10^{ -6}}$&$^{+  2.90\times 10^{ -7}}_{ -2.70\times 10^{ -7}}$ \\
        Cl$^{+}$ (CELs)                         & ${  3.36\times 10^{ -9}}$&$^{+  1.09\times 10^{ -9}}_{ -7.70\times 10^{-10}}$ \\
         Cl$^{2+}$ (CELs)                         & ${  8.48\times 10^{ -8}}$&$^{+  6.80\times 10^{ -9}}_{ -6.30\times 10^{ -9}}$ \\
        Cl$^{3+}$ (CELs)                         & ${  1.54\times 10^{ -8}}$&$^{+  1.10\times 10^{ -9}}_{ -1.00\times 10^{ -9}}$ \\
         \hline 
\end{tabular}
\end{table}

O$^{2+}$ permits the best determination of the abundance discrepancy, as ORLs and CELs of the ion are both present in optical spectra.  68 O~\textsc{ii} recombination lines are detected, together with the three strong [O~\textsc{iii}] CELs at 4363, 4959 and 5007{\AA}.  The abundances derived from the well-detected multiplets V1, V2, V10, V12, and V19 agree very well with each other; lines from multiplets V5, V20, V25, V28 and a number of 3d--4f transitions are detected, but imply much higher abundances and may be affected by noise or bleeding from strong lines in adjacent orders.  As such, the total O$^{2+}$ abundance is derived based only on the 5 well detected multiplets.  The abundance for each multiplet was determined from a flux-weighted average of the detected components, and the overall abundance calculated as the average of the multiplet abundances.  The O$^{2+}$/H$^+$ abundance thus derived is (6.0$\pm$0.3$)\times$10$^{-4}$, while that derived from the strong CELs is (3.4$\pm$0.2$)\times$10$^{-4}$, giving an abundance discrepancy factor (adf; the ratio of ORL abundance to CEL abundance) of 1.75$\pm$0.15.

In the case of oxygen, O$^{2+}$ ORLs and CELs were detected while for O$^+$ only CELs were detected.  The He$^{2+}$/H$^+$ ratio implies that there should be negligible O$^{3+}$, and so the CEL total abundance is simply the sum of O$^+$ and O$^{2+}$.  For ORLs, it was assumed that the ionisation structure is the same, and that the O$^+$/O ratio from CELs can be used to correct for it.  Thus, the abundance discrepancy factors for O$^{2+}$ and for O are the same.

Lines of Ne$^{2+}$ from both recombination and collisional excitation were also both detected in the spectra.  The ORLs give abundances consistent with each other, yielding a value of (1.7$\pm$0.2$)\times$10$^{-4}$.  The CELs give (0.9$\pm$0.05$)\times$10$^{-4}$, resulting a discrepancy factor of 1.9$\pm$0.3, consistent with that derived for O$^{2+}$.

Optical spectra contain CELs only of N$^+$, but ORLs of both N$^{2+}$ and N$^{3+}$.  Most of the N is in the form of N$^{2+}$, and so the comparison of abundances relies on the large and uncertain correction of the CEL abundance for the unseen ions. \cite{delgado-inglada14} proposed a new ICF scheme to compute N abundances; however, the differences with the classical ICF approach (N/O=N$^+$/O$^+$) can be very large, especially for relatively low-excitation PNe, as is the case of IC\,4776.  Using the newer ICF, we obtain N/H = ${  1.67\times 10^{ -4}}^{+  3.00\times 10^{ -5}}_{ -3.60\times 10^{ -5}}$, while the classical approach gives an N abundance $\sim$0.4 dex lower. \cite{delgado-inglada15} found that the \cite{delgado-inglada14} ICF for N showed a correlation with both He abundance and degree of ionisation and, as such, favoured the classical approach. As the ICF for N from \citet{delgado-inglada14} is strongly dependent on whether the PN is radiation or matter-bounded (providing a more realistic correction for matter-bounded PNe; Delgado-Inglada, private communication), we used the classical ICF approach to derive N abundances for IC~4776. The derived abundance discrepancy factor for N is 2.9, with statistical uncertainties of $^{+0.7}_{-0.4}$, though with systematic uncertainties probably much larger (especially given the uncertainty in the choice of ICF). The net result is that, whatever ICF we chose, the N/O ratio is relatively low (-0.75 $<$ log (N/O) $<$ -0.35).

Without UV spectra, we cannot derive an abundance discrepancy factor for carbon.  However, studies have generally found that ionic and elemental ratios from ORLs are consistent with those derived from CELs \citep{wangliu07, delgadoingladarodriguez14}.  The C/O ratio derived from recombination lines is 0.3, atypically low for a planetary nebula (which normally present values around unity\footnote{ \citet{wesson05} found an average C/O ratio of 0.85 in a sample of nebulae outside the solar circle, while \citet{kingsburgh94} find a value of 1.15 in a sample interior to the solar circle - both measured from CELs.}). 

Very recently, \citet{juandedios17} have demonstrated the critical role of atomic data uncertainties in the determination of chemical abundances. They showed that the transition probabilities of the commonly used density diagnostic lines of S$^+$, O$^+$, Cl$^{2+}$ and Ar$^{3+}$ as well as the collision strengths of Ar$^{3+}$ are responsible for most of the uncertainty in the derived total abundances, especially in the high-density (log ($n_e$) $> 4.0$) regime. Given that the computed density for IC~4776 is relatively high, especially for the medium ionisation zone, it is important to note that the uncertainties derived for the total nebular abundances are almost certainly underestimated.
However, the consistency between the adfs derived for O$^{2+}$ and Ne$^{2+}$ provides a strong indication of the validity of the results.

 \begin{table}
	\centering
	\caption{Total nebular abundances, relative to H, in IC~4776.}
	\label{tab:abund}
	\begin{tabular}{rrl}
\hline
        Element & \multicolumn{2}{c}{$\frac{\mathrm{X}}{\mathrm{H}}$}\\
        \hline
        He (ORLs)                                & ${  0.10}$&$\pm{  2.00\times 10^{ -3}}$ \\ 
        C (ORLs)                                 & ${  1.94\times 10^{ -4}}$&$\pm{  4.00\times 10^{ -6}}$ \\ 
		N (ORLs)                                 & ${  1.85\times 10^{ -4}}$&$\pm{  1.10\times 10^{ -5}}$ \\
        N (CELs)                                 & ${ 6.38\times 10^{ -5}}$&$^{+  9.80\times 10^{ -6}}_{ -1.24\times 10^{ -5}}$ \\
        adf (N)                                  & ${  2.92}$&$^{+  0.73}_{ -0.43}$ \\
        O (ORLs)                                 & ${  6.50\times 10^{ -4}}$&$^{+  5.10\times 10^{ -5}}_{ -3.80\times 10^{ -5}}$ \\
        O$^{}$ (CELs)                            & ${  3.71\times 10^{ -4}}$&$^{+  3.20\times 10^{ -5}}_{ -2.40\times 10^{ -5}}$ \\
        adf (O)                                  & ${  1.75}$&$^{+  0.15}_{ -0.14}$ \\
        Ne (ORLs)                                & ${  1.87\times 10^{ -4}}$&$^{+  2.70\times 10^{ -5}}_{ -2.30\times 10^{ -5}}$ \\
        Ne (CELs)                                & ${  1.04\times 10^{ -4}}$&$^{+  9.00\times 10^{ -6}}_{ -7.00\times 10^{ -6}}$ \\
		adf (Ne)                                 & ${  1.77}$&$^{+  0.26}_{ -0.23}$ \\
        Ar$^{}$ (CELs)                           & ${  2.05\times 10^{ -6}}$&$^{+  1.80\times 10^{ -7}}_{ -2.30\times 10^{ -7}}$ \\
        S$^{}$ (CELs)                            & ${  6.73\times 10^{ -6}}$&$^{+  9.90\times 10^{ -7}}_{ -8.60\times 10^{ -7}}$ \\
        Cl$^{}$ (CELs)                           & ${  1.04\times 10^{ -7}}$&$^{+  8.00\times 10^{ -9}}_{ -7.00\times 10^{ -9}}$ \\ 
        \hline
        \end{tabular}
\end{table}


\section{Central star}
\label{sec:cspn}

\subsection{Observations and data reduction}
\label{sec:cspnobs}

The central star of IC~4776 was observed a total of 10 times using the FORS2 instrument mounted on ESO's VLT-UT1 \citep{FORS}.  1200s exposures were acquired in service mode, randomly distributed throughout the observing period (see Table \ref{tab:RVs} for the exact dates).  For each observation, the same instrumental setup was used, employing a longslit of width 0.5\arcsec{} and the GRIS\_1200B grism resulting in a spectral range of 3700\AA{} $\lesssim\lambda\lesssim$ 5100\AA{} with a spectral resolution of approximately 0.8\AA{}.

All data were bias-subtracted, wavelength-calibrated and optimally-extracted using standard \textsc{starlink} routines \citep{figaro}.

\subsection{Radial velocity variability}
\label{sec:cspnrv}
\begin{figure*}
	\includegraphics[width=\textwidth]{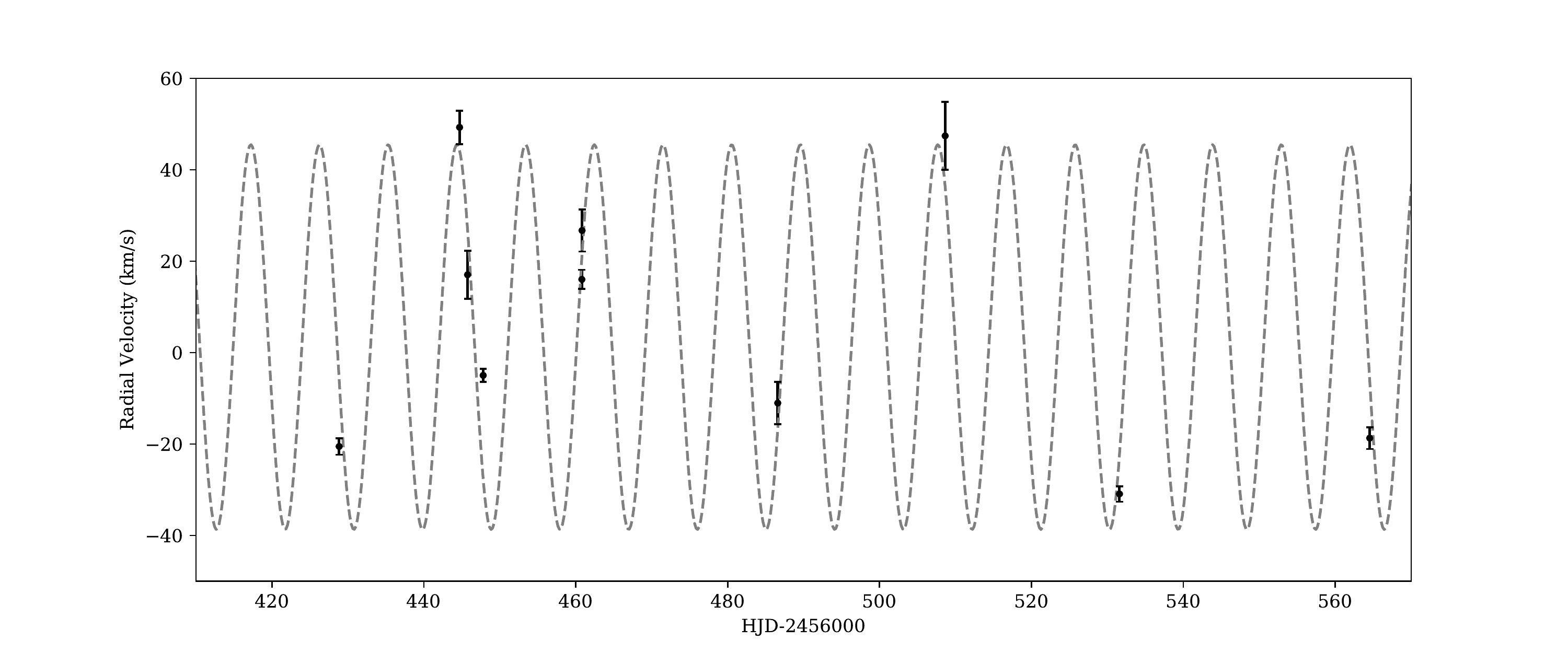}
    \caption{Radial velocity curve of the central star of IC~4776 based on the FORS2 data with a tentative fit of period equal to 9 days. It is important to note that similarly good fits can be made with shorter periods, including periods around 0.3 and 1.2 days.  Further observations are essential to fully constrain the period.}
    \label{fig:RVplot}
\end{figure*}

After the reduction, the spectra were continuum subtracted and aligned using the He~\textsc{i}~$\lambda$4471.48 nebular line, such that the systemic nebular velocity was at 0~\kms{} (thus accounting for both heliocentric velocity variations between observations and small deviations in the wavelength calibration).
The Balmer lines were all found to be severely contaminated by the bright surrounding nebula, essentially leaving only the absorption line of He~\textsc{ii} $\lambda$4541 as a ``clean'' feature for cross-correlation.  The individual spectra were all cross-correlated against a custom mask (namely a flat continuum with a deep, narrow absorption spike at the rest velocity of the He~\textsc{ii} feature) with the resulting radial velocities (c.f. the nebular systemic velocity) listed in Table~\ref{tab:RVs}, and plotted in Figure~\ref{fig:RVplot}. 

\begin{table}
	\centering
	\caption{Radial velocities of the central star of IC~4776 with respect to the nebular systemic velocity}
	\label{tab:RVs}
	\begin{tabular}{lrl}
		\hline
		Heliocentric Julian & \multicolumn{2}{c}{Radial velocity}\\
        Date (days) & \multicolumn{2}{c}{ (\kms{})}\\
		\hline
2456428.84873  &  -20.53 &$\pm$ 1.80 \\
2456444.70375  &  49.28 &$\pm$ 3.66 \\
2456445.75583  &  17.03 &$\pm$ 5.28 \\
2456447.81343  &  -4.98 &$\pm$ 1.42 \\
2456460.80640  &  16.01 &$\pm$ 2.08 \\
2456460.82896  &  26.72 &$\pm$ 4.59 \\
2456486.59737  &  -11.03 &$\pm$ 4.61 \\
2456508.64040  &  47.42 &$\pm$ 7.44 \\
2456531.59275  &  -30.92 &$\pm$ 1.68 \\
2456564.53724  &  -18.70 &$\pm$ 2.37 \\
		\hline
	\end{tabular}
\end{table}

There is a clear variation in the radial velocity with a semi-amplitude of about 30--40 \kms{} strongly indicating that the central star of IC 4776 is a binary system. Unfortunately, with so few radial velocity points, it is not possible to fully constrain the orbital period of the likely binary, with several periodicities and amplitudes providing reasonable fits to the data.  Our current radial velocity data rule out periods much longer than 20 days, but shorter periods are relatively well fit. In Figure~\ref{fig:RVplot}, we show a tentative fit with a period of 9 days to highlight the significant radial velocity variations most likely due to orbital motion, as well as the need for further data.  Interestingly, although the stellar systemic velocity (the zero of the sine curve used to fit the stellar radial velocities, often referred to as $\gamma$) was allowed to vary as part of the fitting procedure, it was found to coincide with the systemic velocity of the nebula (at zero in the plot given that the all radial velocities have been measured with respect to the nebular systemic velocity) within uncertainties ($\Delta v=v_{neb}-\gamma\sim$3~\kms).  This indicates that the plotted fit may indeed be close to the true orbital solution.  However, it must be emphasised that we find similarly good fits for a range of periods, including $\sim$0.3 and $\sim$1.2 days. 

It is important to note that we do not see any clear signs of a secondary component in our spectra.  Previous authors have classified the central star as a \textit{wels} type \citep{tylenda93}, indicating that perhaps some emission lines attributable to the secondary could be present in the spectra.  However, the extremely bright and compact nature of the central region of the nebula can account for the presence of these lines \citep{basurah16}.  We find the same bright, narrow emission lines in our 2D spectra, but find no evidence of an appreciable contribution to their flux from the central star; rather, they originate purely from the compact nebula (although accurate nebular subtraction is a challenge).  Assuming that no such stellar emission lines are present and that the secondary is a main sequence star \citep[the most common scenario;][]{jones17c}, short periods ($\leqslant$1 day) are unlikely given that at short periods a main-sequence star would be expected to show strong emission lines due to irradiation \citep{miszalski11b,jones14a,jones15}.  However, if the secondary is a fainter white dwarf, then short periods cannot be ruled out on the basis of a non-detection of irradiated lines in the spectrum.

Assuming that the nebular inclination derived in Section \ref{sec:nebshape} is reflective of the inclination of the binary \citep[as found in all cases where both inclinations are known; ][]{hillwig16}, then it is possible to place some limits on the possible mass of the secondary.  Further assuming that the amplitude of the fit shown in Figure~\ref{fig:RVplot}, K$_1\sim$40 km~s$^{-1}$, is representative of the true amplitude (a seemingly reasonable assumption given the relatively even distribution of the radial velocity data points), then a 9 day period would imply a mass ratio ($q$=$\frac{M_2}{M_1}$) of around unity.  Longer periods and/or greater radial velocity amplitudes would both imply higher mass ratios and vice versa.  The mass function for the secondary can be written as
\begin{equation}
f(M_2)=\frac{M_2^3 \mathrm{sin}^3 i}{(M_1+M_2)^2}=\frac{P K_1^3}{2\pi\mathrm{G}},
\end{equation}
where $M_1$ and $M_2$ are the primary and secondary masses, $i$ is the inclination of the orbital plane, and $P$ is the orbital period.

\begin{figure}
	\includegraphics[width=\columnwidth]{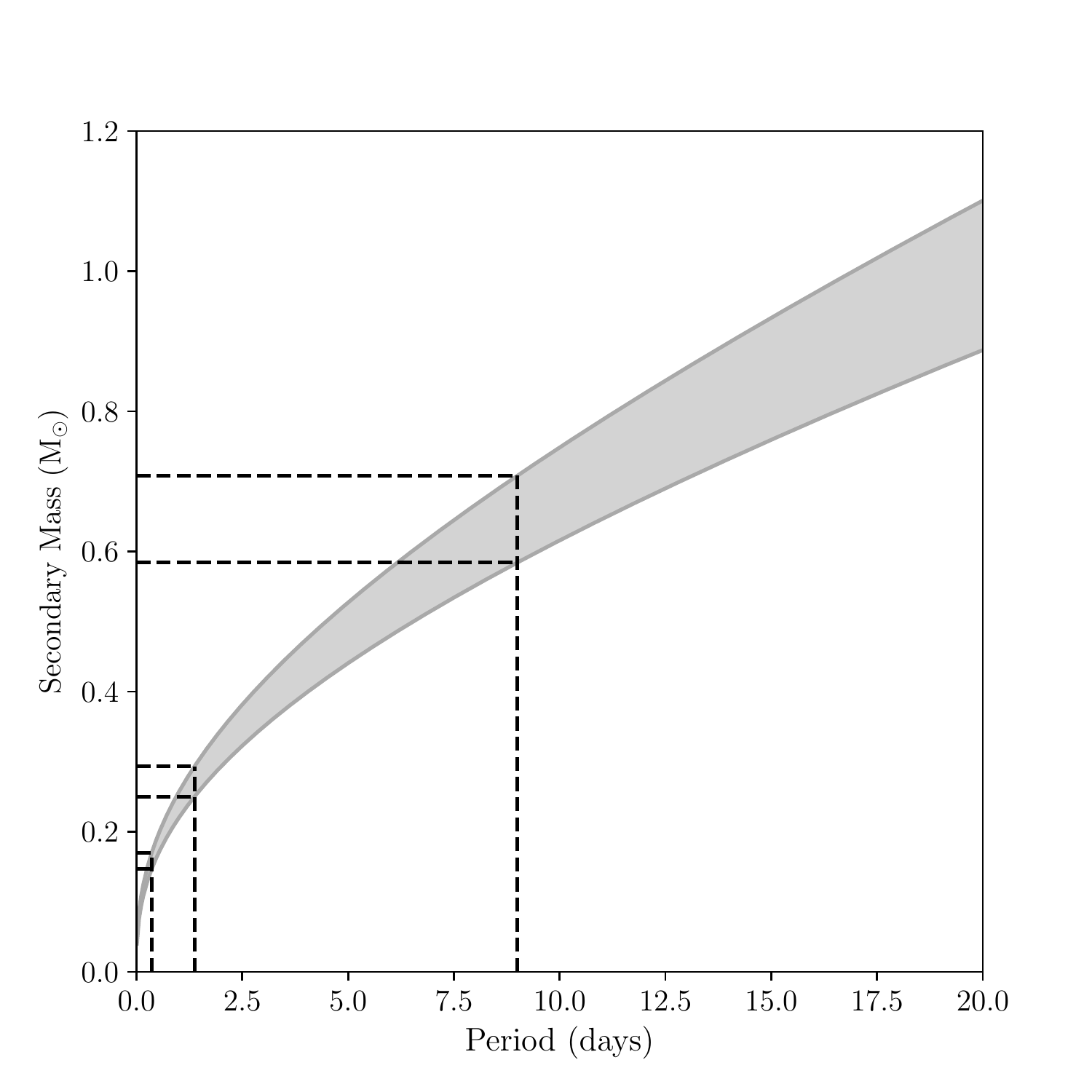}
    \caption{A plot showing the possible secondary masses of IC~4776 (shaded region) as a function of period for the FORS2 radial velocity data. The plot assumes the same radial velocity amplitude as the fit of Figure~\ref{fig:RVplot} and a 0.6M$_\odot$ primary in a circular orbit.  The range of possible masses for a given period is derived assuming that the binary is coplanar with the waist of the nebula, and the range of values for a given period is a reflection of the uncertainty on the nebular inclination.  The periods which offer similarly good fits to the radial velocity data are delineated to highlight the approximate range of possible secondary masses ($\sim 0.1--0.7$ M$_\odot$) For more detail refer to the text.}
    \label{fig:period_mass}
\end{figure}

Taking a relatively standard white dwarf mass for the primary of M$_1$=0.6M$_\odot$ (and maintaining the assumption that K$_1$$\sim$40 km~s$^{-1}$), the mass function can be solved analytically for the secondary mass.  Taking into account the uncertainty on the nebular inclination (based on the spatiokinematical modelling presented in Section \ref{sec:nebshape}), the possible range of secondary masses are plotted as a function of period in Figure~\ref{fig:period_mass}.  Under these assumptions, the tentative 9 day period would imply that the secondary is likely a late-type main sequence star (roughly of spectral type K) or a white dwarf.  For the secondary to be a white dwarf, it would have to have been initially the more massive component of the binary and, as such, would be expected to leave a more massive remnant (at least more massive than the primary remnant which, in this case, was assumed to be 0.6M$_\odot$).  This is a possibility given that the open mass range for the secondary, based on our fits, is approximately 0.6--0.7M$_\odot$.  However, a more massive secondary would probably be required in order for it to have now cooled beyond observability in our spectra.  If the true period is shorter than 9 days (Figure~\ref{fig:period_mass} also shows solutions for the reasonable fits at 0.3 days and 1.2 days) and  then the open mass range would be significantly lower, ruling out an evolved white dwarf companion for the same reason, while a main sequence secondary would become less likely given the aforementioned non-detection of irradiated emission lines originating from the secondary.  However, this effect may be counteracted by the implied low secondary masses (and therefore radii) at shorter periods, as the level of irradiation in such systems is principally a function of the apparent radius of the irradiated star (proportional to the ratio of the secondary radius and the orbital separation).

Greater radial velocity amplitudes than the assumed $\sim$40 km~s$^{-1}$ would open the possibility of a more massive white dwarf companion, but the currently available data provides no indication of such a large amplitude.  Similarly, periods longer than 9 days open the parameter space to the possibility of a more massive, evolved white dwarf companion, however in that case the system would have to have undergone two common-envelope phases (given that the AGB radius of a more massive companion would inevitably have been larger than that of the primary) which almost certainly would result in a very short orbital period \citep{tovmassian10}.  Therefore, on a stellar evolutionary basis, a main sequence secondary (roughly of K-type for a period of 9 day, or late M-type for periods less than 1 day) would seem the more plausible companion, although a white dwarf secondary cannot be entirely ruled out.

\section{Discussion}
\label{sec:discussion}

Population synthesis models predict many post-CE systems with periods of several to a few tens of days \citep[e.g.][]{han95}. However, the known population of post-CE binary central stars of PNe is very sparse in this region \citep{demarco08,jones17c}, with a similar paucity observed in the general white dwarf plus main sequence binary population \citep{nebot11}.  Thus far, it is unclear whether this lack of intermediate period binary central stars represents purely failure of the population synthesis models\footnote{It is important to note that population synthesis models rely on an ad hoc prescription of the common envelope process, often relying on simple parameterisations that have been proven to be less than satisfactory in reproducing observations \citep{demarco11b}.  Hydrodynamic models thus far fall short of being able to make the predictions required for use in population synthesis \citep{ivanova13}, however tend to almost unanimously predict very short post-CE orbitals periods while the few that do predict longer periods seem to be under-resolved \citep[see e.g.][]{passy12}.} or also an observational bias towards the discovery of short period systems.  

Recent observations have indicated that long-term radial velocity monitoring with modern, high-stability spectrographs may hold the key to revealing this missing intermediate-period population, should it indeed exist \citep[e.g.][]{miszalski17}.  Here, we have reported on the discovery of a post-CE binary star at the centre of IC~4776 via such radial velocity monitoring.  While the period of the system cannot be confirmed by the data (several periods ranging from $\sim$0.3--9 days provide reasonable orbital solutions), the detection further supports the hypothesis that a systematic radial velocity survey may indeed reveal the presence of many intermediate period binaries inside PNe.

The central star of IC~4776 was found to be a single-lined binary with a tentative orbital period of $\sim$9 days (though shorter periods cannot be ruled out), where the secondary is most likely a main sequence star (of mass $\sim$0.1--0.7 M$_\odot$).  The system was selected for monitoring based primarily on the presence of precessing jets, believed to result from mass transfer in a central binary system.  Such mass transfer has been hypothesized to occur either prior to or immediately after the CE phase \citep{corradi11,miszalski13b,boffin12b,tocknell14}, but may also help to prevent in-spiral during the CE as part of a grazing envelope evolution \citep[GEE;][]{soker15,shiber17}.  Such a GEE may lead to the preferential formation of wide binaries, but the range of possible periods derived for IC~4776 does not extend to long enough periods to necessitate a GEE to explain its formation (though nor does it rule out such an evolution).

An empirical analysis of the chemical abundances in IC~4776 based on \'echelle spectra covering the entire optical range indicates a low N/O ratio as well as a low C/O ratio (both of which are $\sim$0.3).  This may be indicative that interaction with the companion cut short the AGB evolution of the nebular progenitor \citep{demarco09,jones14a}.

In PNe, there is a long standing discrepancy between chemical abundances derived using ORLs and those derived using CELs, which has often been attributed to the existence of a second phase of cold hydrogen deficient material within the normal nebular gas phase \citep{liu00,tsamis03,zhang04,wesson05}.  The discrepancy is typically a factor of 2--3 but exceeds a factor of 5 in about 20\% of nebulae. Recent observations have suggested that high abundance discrepancies may be strongly correlated with central stars that have undergone a CE evolution \citep{corradi15,jones16b,wesson17}. However, our chemical analysis reveals the adf of IC~4776 to be particularly low at $\sim$2, which is not just low for a PN with a binary central star, but for PNe in general.  Interestingly, the only other post-CE system known to reside in a low-adf nebula is an intermediate-period system \citep[NGC~5189 with a period of 4 days;][]{garcia-rojas13,manick15} which, if the period of IC~4776 is a long as 9 days, may imply a connection between intermediate period binaries and low nebular adfs.  Furthermore, NGC~5189 is host to a [WR] type primary \citep{manick15}, as has been claimed for the central star of IC~4776 \citep{tylenda93}.  In total, this means three [WR]-type central stars are known to reside in binaries, with the two previous detections (NGC~5189 and PM~1-23) both having orbital periods longer than one day\footnote{The orbital period of the [WR] central star of PM~1-23 is given as 0.6 days by \cite{hajduk10}, but further observations have shown that this is, in fact, half the orbital period \citep[][Miszalski et al., in prep.]{manick15}.}.  

The connection between intermediate periods, low adfs and [WR] central stars is, thus far, tenuous, given the extremely small number statistics involved, particularly given that we are unable to derive a definitive period for the central star of IC~4776.  However, there may perhaps be reasonable physical grounds for such correlations. \cite{manick15} suggest that the strong winds observed in [WR] stars may help to prevent spiral-in, via pre-CE wind interaction, leading to longer post-CE periods\footnote{A similar scenario has been suggested for the intermediate-period binary central star of NGC~2346 \citep[V651 Mon, the period of which is 16 days;][]{soker02b}, but there is no indication that its central star is of the [WR]-type.}.  

Should high adfs be the result of a nova-like eruption which ejects a second, chemically-enriched gas phase into the nebula \citep{corradi15,jones16b,garcia-rojas16}, then it is not unreasonable to expect that central stars with little hydrogen on their surfaces may be less likely to experience such outbursts (i.e.\ without hydrogen to burn, they do not undergo an eruption).  However, one formation mechanism for such [WR] stars is to undergo some form of late thermal pulse which depletes their outer layers in hydrogen, whilst also providing an explanation for the presence of hydrogen-poor ejecta in their host nebulae  \citep{ercolano04}.  Indeed, previously connections were made between the abundance discrepancy problem and hydrogen-depleted central stars \citep{ercolano04}, particularly in the case of born-again central stars which are found to show some of the highest discrepancies \citep{wesson03,wesson08b}.  However, more recent studies have shown that in many cases PNe with [WR] central stars present with rather modest abundance discrepancies \citep{garcia-rojas13}.  

Assuming that high adfs are the result of late ejection of low-metallicity material, it might be expected that longer period systems do not experience any post-CE mass transfer that could lead to an outburst.  There is little evidence of such post-CE mass transfer in other high adf PNe, however, in NGC~6337, jets are observed to have been launched after the ejection of the CE \citep[perhaps as a result of post-CE mass transfer;][]{garcia-diaz09} while the nebular adf is observed to be highly elevated \cite[$\sim$30;][]{wesson17}.  Furthermore, the possibility remains open for other systems based on the observation of near-Roche-lobe-filling, post-CE secondary stars \citep{jones15}.  

Unfortunately, the adf of the only other PN known to host a [WR]-binary central star (PM~1-23) is not known, nor are the adfs of other PNe known to host long- and intermediate-period post-CE central stars (e.g.\ LTNF~1, Sp~1, 2MASS~J19310888+4324577, MPA~J1508$-$6455, NGC~2346, NGC~1360).  Thus, making further assessment of the possible relationships between periodicity, abundance pattern and central star type currently impossible.

In conclusion, in spite of the laborious nature of the observations \citep{demarco04}, we encourage further radial velocity studies of the central stars of planetary nebulae in search of the missing population of intermediate-period post-CE binaries.  As highlighted here, the pre-selection of targets based on morphology may be the most effective means of increasing the hit-rate for detections. Whilst the possibility of a connection between [WR]-type central stars and intermediate periods is intriguing (as well as a possible link between [WR]-type and low adfs), a more rigorous survey of [WR] central stars (and their host nebulae) is required before any strong conclusions can be drawn.  Furthermore, a connection between intermediate-period central stars and low adfs is on a similarly uncertain footing, and abundance studies of the known sample of PNe known to host binary central stars (of all periodicities) is strongly encouraged.  Further observations of the central star of IC~4776 are essential in order to full constrain its period, and begin to place these nascent relationships on a stronger footing.

\section*{Acknowledgements}
We are grateful to the referee, Orsola De Marco, for helpful comments and suggestions that further improved the paper. We would like to thank Tom Marsh for the use of his \textsc{molly} software package, and An\'ibal Garc\'ia-Hernandez for the use of his UVES data. We thank Gloria Delgado-Inglada for fruitful discussions.
Based on data obtained at the European Southern Observatory, Chile, under proposal numbers 	097.D-0037, 091.D-0673, 089.D-0429, 087.D-0446. This research made use of Astropy, a community-developed core Python package for Astronomy \citep{astropy}.
PS acknowledges the ING Support and Research Studentship, and thanks the Polish National Center for Science (NCN) for support through grant 2015/18/A/ST9/00578. J.~G-R acknowledges support from Severo Ochoa Excellence Program (SEV-2015-0548) Advanced Postdoctoral Fellowship.




\bibliographystyle{mnras}
\bibliography{literature} 

\appendix

\section{UVES spectrum of IC~4776}
Figure \ref{fig:alfaneatfull} shows the full UVES spectrum of IC~4776 referred to in Section \ref{sec:neb}, along with the \textsc{alfa} fit over-plotted.  The observed and dereddened fluxes derived from the \textsc{alfa} fit are presented in Table \ref{tab:fluxes}.

 \begin{table*}
	\centering
	\caption{Observed, $F(\lambda)$, and dereddened, $I(\lambda)$, nebular emission line fluxes of IC~4776 relative to $I(\mathrm{H}\beta) = 100$ ($\equiv F(\mathrm{H}\beta) = 103.992$). Asterisks denote that the line is detected as a blend with the previous line in the table, and the flux listed is the total flux of the unresolved lines. The upper and lower term columns show the upper and lower levels of the atomic transition producing the line, while g$_1$ and g$_2$ are the degeneracy factors of the two levels.}
	\label{tab:fluxes}
\begin{tabular}{rlrllllllll}
\hline
  \hline
   $ \lambda $ (\AA{}) & \multicolumn{2}{c}{$F \left( \lambda \right) $ }& \multicolumn{2}{c}{$I \left( \lambda \right) $} & Ion & Multiplet & Lower term & Upper term & g$_1$ & g$_2$\\
 \hline 
3312.32 &    0.037& $\pm$   0.007&   0.029& $\pm$   0.008 & O~{\sc iii}      & V3         & 3s 3P*     & 3p 3S      &          3 &        3    \\
3322.53 &    0.050& $\pm$   0.015&   0.071& $\pm$   0.018 & [Fe~{\sc iii}]   &            & 5D         & 7S         &          9 &        7    \\
3334.84 &    0.100& $\pm$   0.012&   0.122& $\pm$   0.015 & Ne~{\sc ii}      & V2         & 3s 4P      & 3p 4D*     &          6 &        8    \\
3342.50 &    0.171& $\pm$   0.020&   0.173& $^{  +0.026}_{  -0.025}$ & [Ne~{\sc iii}]   &            & 2p4 1D     & 2p4 1S     &          5 &        1    \\
3353.17 &    0.036& $\pm$   0.011&   0.060& $\pm$   0.013 & [Cl~{\sc iii}]   &            & 3p3 4S*    & 3p3 2P*    &          4 &        2    \\
3354.55 &    0.095& $\pm$   0.014&   0.127& $\pm$   0.017 & He~{\sc i}       & V8         & 2s 1S      & 7p 1P*     &          1 &        3    \\
3355.02 &    0.035& $\pm$   0.014&   0.060& $^{  +0.016}_{  -0.017}$ & Ne~{\sc ii}      & V2         & 3s 4P      & 3p 4D*     &          4 &        6    \\
\end{tabular}
\end{table*}

\begin{table*}
\contcaption{Observed and dereddened emission line fluxes from IC~4776}
\begin{tabular}{rlrllllllll}
\hline
$ \lambda $ (\AA{}) & \multicolumn{2}{c}{$F \left( \lambda \right) $ }& \multicolumn{2}{c}{$I \left( \lambda \right) $} & Ion & Multiplet & Lower term & Upper term & g$_1$ & g$_2$\\
\hline
3646.50 &    0.195& $\pm$   0.006&   0.222& $\pm$   0.008 & Balmer Jump+                                                                       \\
3819.62 &    1.154& $\pm$   0.027&   1.323& $\pm$   0.040 & He~{\sc i}       & V22        & 2p 3P*     & 6d 3D      &          9 &       15    \\
3835.39 &    6.048& $\pm$   0.066&   6.865& $^{  +0.142}_{  -0.152}$ & H~{\sc i}        & H9         & 2p+ 2P*    & 9d+ 2D     &          8 &        *    \\
3856.02 &    0.082& $\pm$   0.008&   0.093& $\pm$   0.002 & S~{\sc iii}      & V12        & 3p2 2D     & 4p 2P*     &          6 &        4    \\
3856.13 &  *     &             & *     &         & O~{\sc ii}       & V12        & 3p 4D*     & 3d 4D      &          4 &        2    \\
3862.60 &    0.064& $\pm$   0.005&   0.079& $^{  +0.006}_{  -0.007}$ & Si~{\sc ii}      & V1         & 3p2 2D     & 4p 2P*     &          4 &        2    \\
3868.75 &   70.487& $\pm$   2.963&  77.200& $\pm$   3.600 & [Ne~{\sc iii}]   & F1         & 2p4 3P     & 2p4 1D     &          5 &        5    \\
\end{tabular}
\end{table*}

\begin{table*}
\contcaption{Observed and dereddened emission line fluxes from IC~4776}
\begin{tabular}{rlrllllllll}
\hline
$ \lambda $ (\AA{}) & \multicolumn{2}{c}{$F \left( \lambda \right) $ }& \multicolumn{2}{c}{$I \left( \lambda \right) $} & Ion & Multiplet & Lower term & Upper term & g$_1$ & g$_2$\\
 \hline
4291.25 &    0.011& $\pm$   0.002&   0.011& $\pm$   0.002 & O~{\sc ii}       & V55        & 3d 4P      & 4f G3*     &          6 &        8    \\
4292.16 &    0.007& $\pm$   0.002&   0.008& $\pm$   0.001 & C~{\sc ii}       &            & 4f 2F*     & 10g 2G     &         14 &       18    \\
4292.21 &  *     &             & *     &         & O~{\sc ii}       & V78c       & 3d 2F      & 4f F2*     &          6 &        6    \\
4294.78 &    0.019& $\pm$   0.002&   0.020& $\pm$   0.001 & O~{\sc ii}       & V53b       & 3d 4P      & 4f D2*     &          4 &        6    \\
4294.92 &  *     &             & *     &         & O~{\sc ii}       & V53b       & 3d 4P      & 4f D2*     &          4 &        4    \\
4303.82 &    0.027& $\pm$   0.002&   0.027& $\pm$   0.002 & O~{\sc ii}       & V53a       & 3d 4P      & 4f D3*     &          6 &        8    \\
4317.14 &    0.050& $\pm$   0.008&   0.046& $\pm$   0.008 & O~{\sc ii}       & V2         & 3s 4P      & 3p 4P*     &          2 &        4    \\
\end{tabular}
\end{table*}

\begin{table*}
\contcaption{Observed and dereddened emission line fluxes from IC~4776}
\begin{tabular}{rlrllllllll}
\hline
$ \lambda $ (\AA{}) & \multicolumn{2}{c}{$F \left( \lambda \right) $ }& \multicolumn{2}{c}{$I \left( \lambda \right) $} & Ion & Multiplet & Lower term & Upper term & g$_1$ & g$_2$\\
 \hline
4685.68 &    0.018& $\pm$   0.006&   0.020& $\pm$   0.006 & He~{\sc ii}      & 3.4        & 3d+ 2D     & 4f+ 2F*    &         18 &       32    \\
4696.35 &    0.008& $\pm$   0.001&   0.007& $\pm$   0.001 & O~{\sc ii}       & V1         & 3s 4P      & 3p 4D*     &          6 &        4    \\
4699.22 &    0.012& $\pm$   0.002&   0.015& $\pm$   0.002 & O~{\sc ii}       & V25        & 3p 2D*     & 3d 2F      &          4 &        6    \\
4701.62 &    0.359& $\pm$   0.022&   0.335& $\pm$   0.022 & [Fe~{\sc iii}]   & F3         & 3d6 5D     & 3d6 3F2    &          7 &        7    \\
4711.37 &    0.158& $\pm$   0.005&   0.158& $\pm$   0.005 & [Ar~{\sc iv}]    & F1         & 3p3 4S*    & 3p3 2D*    &          4 &        6    \\
4713.17 &    0.744& $\pm$   0.014&   0.760& $\pm$   0.015 & He~{\sc i}       & V12        & 2p 3P*     & 4s 3S      &          9 &        3    \\
4733.91 &    0.152& $\pm$   0.007&   0.152& $\pm$   0.008 & [Fe~{\sc iii}]   & F3         & 3d6 5D     & 3d6 3F2    &          5 &        5    \\
\end{tabular}
\end{table*}

\begin{table*}
\contcaption{Observed and dereddened emission line fluxes from IC~4776}
\begin{tabular}{rlrllllllll}
\hline
$ \lambda $ (\AA{}) & \multicolumn{2}{c}{$F \left( \lambda \right) $ }& \multicolumn{2}{c}{$I \left( \lambda \right) $} & Ion & Multiplet & Lower term & Upper term & g$_1$ & g$_2$\\
 \hline
7065.25 &   11.038& $\pm$   0.254&   8.808& $\pm$   0.340 & He~{\sc i}       & V10        & 2p 3P*     & 3s 3S      &          9 &        3    \\
7135.80 &   17.979& $\pm$   0.671&  14.576& $\pm$   0.716 & [Ar~{\sc iii}]   & F1         & 3p4 3P     & 3p4 1D     &          5 &        5    \\
7160.56 &    0.033& $\pm$   0.001&   0.026& $\pm$   0.001 & He~{\sc i}       &            & 3s 3S      & 10p 3P*    &          3 &        9    \\
7231.32 &    0.029& $\pm$   0.003&   0.026& $\pm$   0.003 & C~{\sc ii}       & V3         & 3p 2P*     & 3d 2D      &          2 &        4    \\
7236.19 &    0.061& $\pm$   0.004&   0.049& $\pm$   0.002 & C~{\sc ii}       & V3         & 3p 2P*     & 3d 2D      &          4 &        6    \\
7236.42 &  *     &             & *     &         & C~{\sc ii}       & V3         & 3p 2P*     & 3d 2D      &          4 &        6    \\
7237.17 &    0.024& $\pm$   0.007&   0.019& $\pm$   0.001 & C~{\sc ii}       & V3         & 3p 2P*     & 3d 2D      &          4 &        4    \\
\end{tabular}
\end{table*}

\begin{table*}
\contcaption{Observed and dereddened emission line fluxes from IC~4776}
\begin{tabular}{rlrllllllll}
\hline
$ \lambda $ (\AA{}) & \multicolumn{2}{c}{$F \left( \lambda \right) $ }& \multicolumn{2}{c}{$I \left( \lambda \right) $} & Ion & Multiplet & Lower term & Upper term & g$_1$ & g$_2$\\
\hline
8680.28 &    0.029& $\pm$   0.002&   0.022& $\pm$   0.001 & N~{\sc i}        & V1         & 3s  4P     & 3p  4D*    &          6 &        8    \\
8680.53 &  *     &             & *     &         & Ca~{\sc iii}     &            &5d 2$<$4$>$*&5f 2$<$3$>$ &          9 &        7    \\
8686.15 &    0.025& $\pm$   0.008&   0.018& $\pm$   0.006 & N~{\sc i}        & V1         & 3s  4P     & 3p  4D*    &          2 &        4    \\
8703.87 &    0.013& $\pm$   0.005&   0.012& $\pm$   0.004 & [Ni~{\sc ii}]    &            & (3F)4s 2F  & (1D)4s 2D  &          8 &        6    \\
8728.90 &    0.021& $\pm$   0.005&   0.012& $\pm$   0.004 & [Fe~{\sc iii}]   &            & 3P4        & 3D         &          5 &        7    \\
8733.43 &    0.056& $\pm$   0.002&   0.043& $^{  +0.002}_{  -0.003}$ & He~{\sc i}       & 3.12       & 3d 3D      & 12f 3F*    &         15 &             \\
8739.97 &    0.003& $\pm$   0.001&   0.002& $\pm$   0.001 & He~{\sc i}       &            & 3d 3D      & 12p 3P*    &          5 &        3    \\
\hline
 \end{tabular}
\end{table*}

\begin{landscape}
 \begin{figure}
     \centering
   \includegraphics[height=0.95\textwidth]{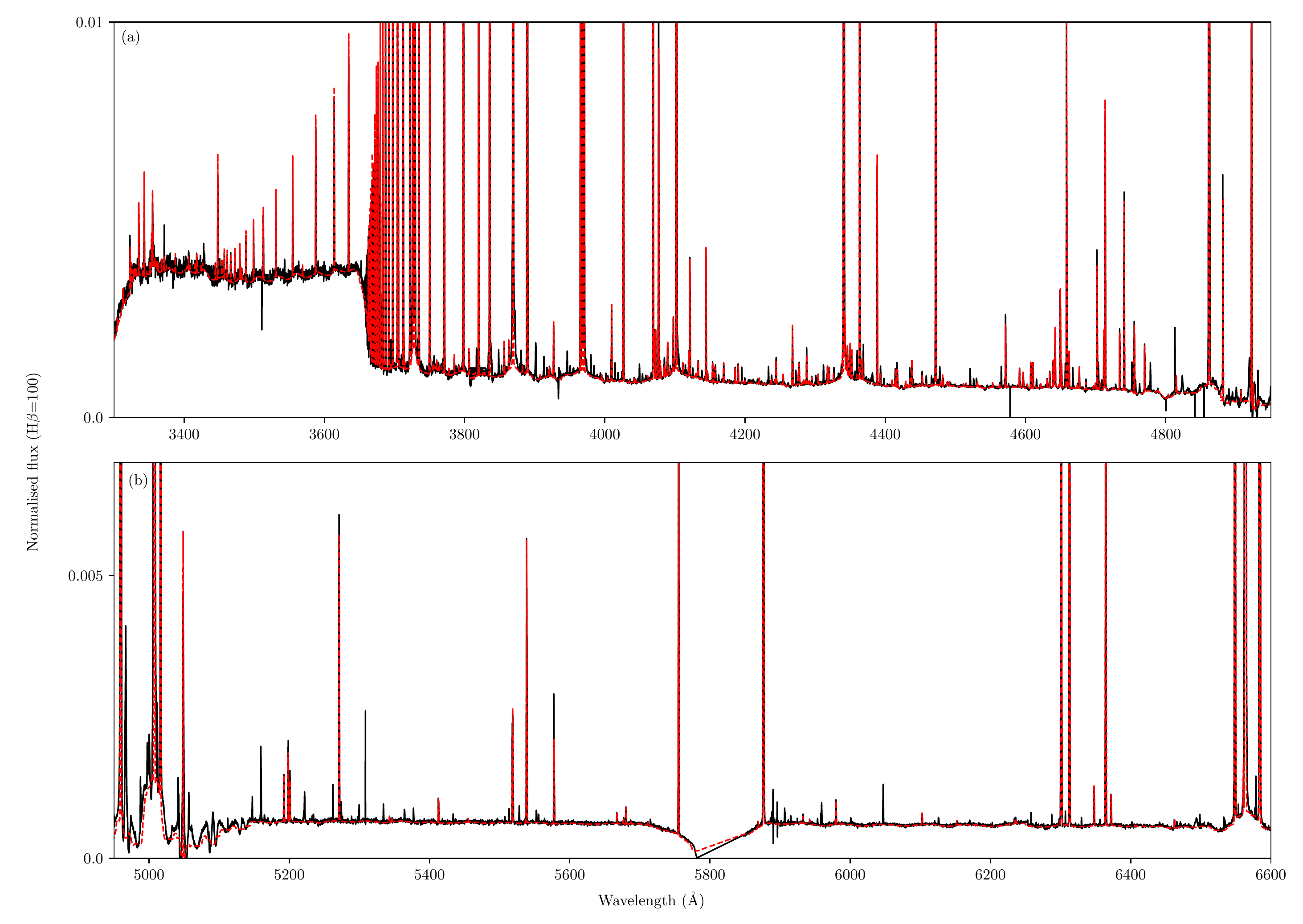}
   \caption{The \textsc{alfa} spectral fit (red-dashed) overlaid on top of the combined, observed UVES spectra (black-solid).}
   \label{fig:alfaneatfull}
\end{figure} 
\end{landscape}

\begin{landscape}
 \begin{figure}
     \centering
   \includegraphics[height=0.95\textwidth]{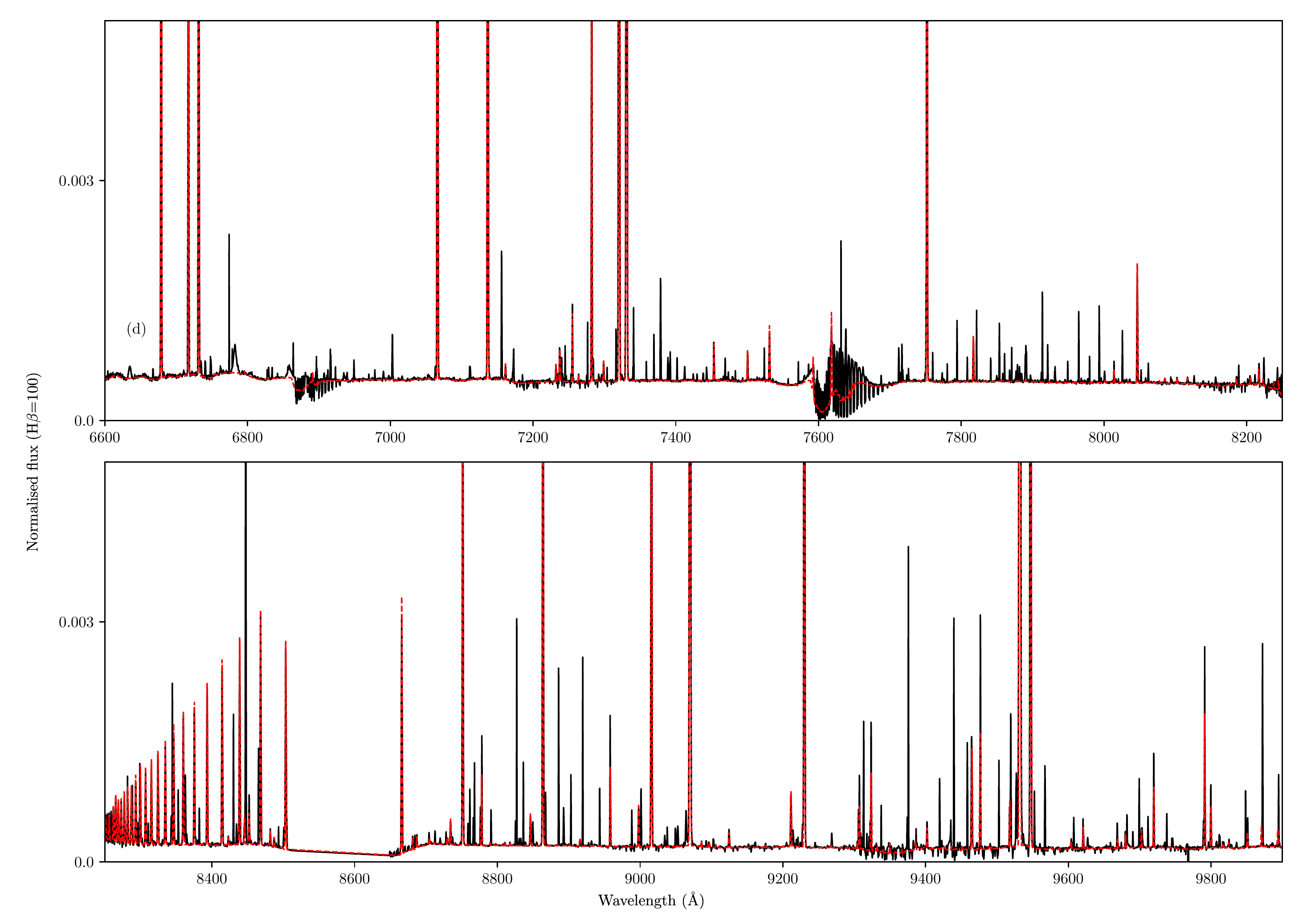}
   \contcaption{The \textsc{alfa} spectral fit (red-dashed) overlaid on top of the combined, observed UVES spectra (black-solid).}
\end{figure} 
\end{landscape}

\bsp	
\label{lastpage}
\end{document}